\documentclass[11pt,a4paper]{article}
\pdfoutput=1
\usepackage[utf8]{inputenc}
\usepackage[english]{babel}

\usepackage{extarrows}
\usepackage{amsmath,bm}
\usepackage{amsfonts}
\usepackage{amssymb}
\usepackage{graphicx}
\usepackage{fourier}

\usepackage{empheq}

\usepackage{caption}
\usepackage{subcaption}
\usepackage{float}
\usepackage{appendix}
\usepackage{todonotes}
\usepackage{authblk}

\numberwithin{equation}{section}

\usepackage[hidelinks]{hyperref}

\newcommand\nin{\noindent}
\newcommand\nn{\nonumber}
\newcommand\be{\begin{equation}}
\newcommand\ee{\end{equation}}
\newcommand\ba{\begin{eqnarray}}    
\newcommand\ea{\end{eqnarray}}      

\title{The Gravity Dual of Real-Time CFT at Finite Temperature}
\author{Marcelo Botta-Cantcheff}
\author{Pedro J. Mart\'inez}
\author{Guillermo A. Silva}
\affil{{\it  Instituto de F\'\i sica de La Plata, CCT La Plata - CONICET \& 

Departamento de F\'\i sica - Universidad Nacional de La Plata 

C.C. 67, 1900 La Plata, Argentina}

~

{\tt E-mail:} botta,martinezp,silva@fisica.unlp.edu.ar}

\usepackage[left=2cm,right=2cm,top=2cm,bottom=2cm]{geometry}

\begin{document}
\maketitle
\begin{abstract}
We present a spherically symmetric aAdS gravity solution with Schwinger-Keldysh boundary condition dual to a CFT at finite temperature defined on a complex time contour.
The geometry is built by gluing the exterior of a two-sided AdS Black Hole, the (aAdS) Einstein-Rosen wormhole, with two  Euclidean black hole halves. These pieces are interpreted as the gravity duals of the two Euclidean $\beta/2$ segments in the SK path, each coinciding with a Hartle-Hawking-Maldacena (TFD) vacuum state, while the Lorentzian regions naturally describes the real-time evolution of the TFD doubled system. 

Within the context of Skenderis and van Rees real-time holographic prescription, the new solution should be compared to the Thermal AdS spacetime since both contribute to the gravitational path integral. In this framework, we compute the time ordered 2-pt functions of scalar CFT operators via a non-back-reacting Klein-Gordon field for both backgrounds and confront the results. When solving for the field we find that the gluing leads to a geometric realization of the Unruh trick via a completely holographic prescription. Interesting observations follow from $\langle {\cal O}_L{\cal O}_R\rangle$, which capture details of the entanglement of the (ground) state and the connectivity of the spacetime.
\end{abstract}
\newpage
\tableofcontents

\section{Introduction}

The study of the gravity/gauge correspondence at finite temperature was initiated by Witten in \cite{haw-page2}, where  the CFT was formulated in periodic imaginary time, i.e.  the Matsubara formalism. In this setup, the conformal field theory (CFT) is formulated on the $S^{d-1} \times S^1_\beta$ boundary. Since the work of Hawking-Page \cite{haw-page1}, two aAdS gravity solutions are known to fulfill the boundary conditions:  the so-called Thermal AdS and the Euclidean AdS black hole, which dominate in the low temperature and high temperature limit respectively. 
 
However, a real time extension of the formalism  is needed for the study of non-equilibrium and finite temperature dynamical processes. The initial steps in this direction started with \cite{herzog} (see also \cite{son}) where, with a Schwinger-Keldysh perspective, the CFT finite temperature propagator matrix elements were reproduced using the maximally extended AdS-BH geometry. Nevertheless, the procedure involved imposing infalling boundary conditions at the horizon, which contrasts with the holographic viewpoint. 

In \cite{SvRC} (see \cite{balas,marolf} for previous work), a Lorentzian formulation of the correspondence was presented. By gluing Euclidean and Lorentzian regions, the prescription relied only on boundary data without resorting to boundary conditions inside the bulk. Within this setup, the Thermal AdS real time extension was easily built \cite{SvRL}. 
The high temperature CFT matrix elements were re-obtained, but at the expense of requiring two copies of the maximally extended AdS black hole \cite{SvRL,Balt}. 

The Skenderis-van Rees (SvR) prescription \cite{SvRC,SvRL}, provides the natural framework to study gravity duals to Schwinger-Keldysh (SK) closed paths \cite{Schwinger,Keldysh,UmezawaAFT, DasTFD}. Among the possible closed path in the complex $t$-plane, Umezawa singled out a particular one which connects the SK and Thermofield Dynamics (TFD) formalisms \cite{UmezawaAFT,TFDUme,Takashi-TFD}. 
It is worth mentioning that already in \cite{herzog} the authors stress that their prescription acquires nice properties for the SK contour highlighted by Umezawa. 
It is this particular path that we will elaborate on in this work.  The TFD interpretation is central in our study and it has already proved useful in the AdS/CFT context in \cite{eternal} where, based in the Hartle Hawking construction \cite{HH}, Maldacena showed that the half Euclidean black hole geometry maps to the TFD vacuum state in the boundary field theory. See \cite{papado} for other works on maximally extended AdS black holes. 

In this work, we present an exact spherically symmetric solution to Einstein gravity dual to the CFT on a SK path mentioned above. 
It is genuinely holographic geometry in the sense that is completely determined by asymptotic boundary data, and in addition, is the natural  real-time extension of the Euclidean AdS-BH by inserting the two-sided exterior of a single black hole. This is in line with Israel's interpretation of the TFD degrees of freedom as being physically realized within the two-sides of BH geometry \cite{Israel}.  The present solution can also be thought of as the real time evolution of the Hartle-Hawking-Maldacena state \cite{eternal} under the TFD Hamiltonian. 

This work is organized as follows: In Sec \ref{SK+CFT}, we review the SvR prescription and its relation to SK thermal paths, i.e. closed time contours. We will describe the CFT theory defined by our contour in the TFD picture and study the predictions for the bulk theory via the duality. We will consider the large $N$ limit and the two contributions to the saddle point approximation of the gravitational path integral, noting that the path admits also a second Thermal-AdS dual which we should compare our results with. In Sec. \ref{DualGeom} we describe in detail the construction of the geometry as well as the boundary conditions for the field inhabiting our geometry. Sec. \ref{BH} will deal with the KG field computations inside our geometry and the gluing procedure. 
For completeness in Sec. \ref{Thermal} we briefly describe the Thermal-AdS geometry. Sec. \ref{Correlators} compares the 2-pt bulk correlators obtained for both geometries with the CFT predictions. We find specially interesting the correlators between the two disconnected CFT's. Finally, Sec. \ref{Conclusions} summarizes the results and possibilities for future work.

\begin{figure}[t]\centering
\begin{subfigure}{0.49\textwidth}\centering
\includegraphics[width=.9\linewidth] {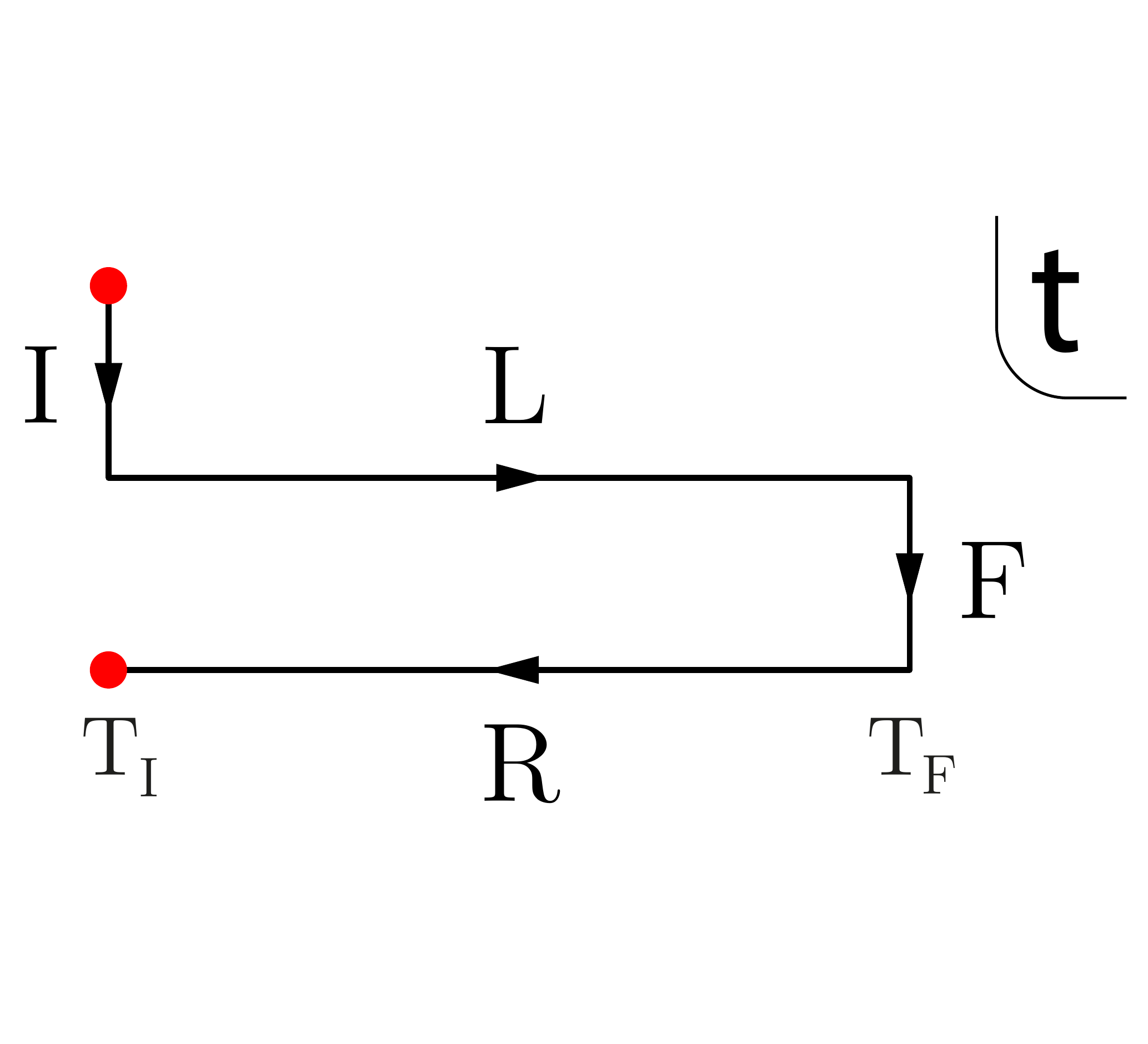}
\caption{}
\end{subfigure}
\begin{subfigure}{0.49\textwidth}\centering
\includegraphics[width=.9\linewidth] {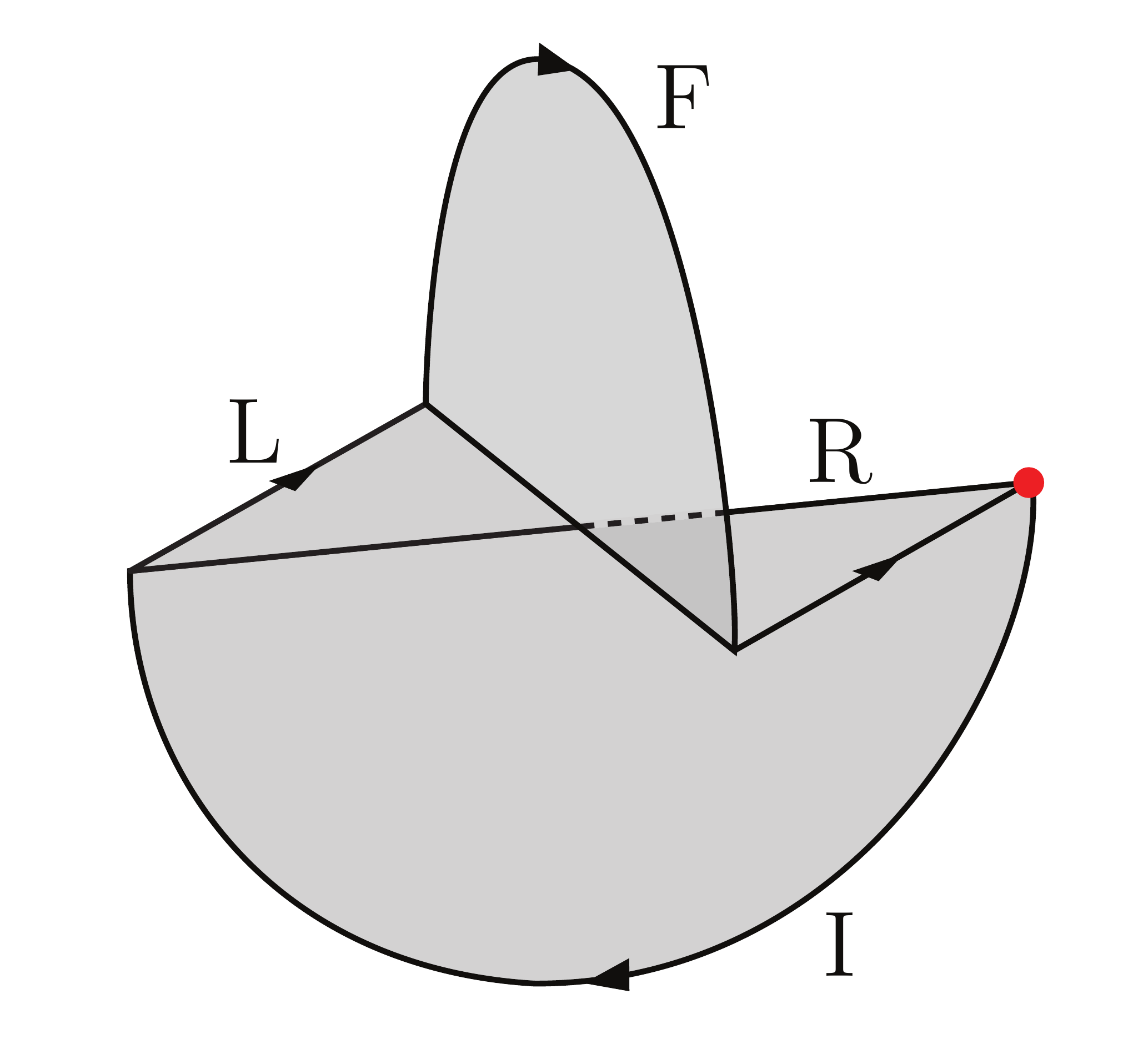}
\caption{}
\end{subfigure}
\caption{(a) Closed Schwinger-Keldysh path in the complex $t$-plane.  The horizontal lines represent real time evolution and the vertical lines give imaginary time evolution. Regions I and II have identical lengths equal to $\beta/2$.  
The red dots are identified, resulting in a closed path.
(b) Dual bulk geometry filling the path on the left. The semicircular pieces represent the Euclidean sections, while the horizontal plane depicts the two sided AdS BH exterior represented as triangular wedges L and R. The arrows along the boundary display the path ordering shown in the left figure. The red dot from the path is also represented. Angular coordinates have been suppressed. }
\label{Fig:Camino}
\end{figure}

\section{Holography for a closed time-contour and TFD}
\label{SK+CFT}
This section is devoted to elaborate on the SvR prescription when considering a conformal field theory defined on a closed time contour in the complex plane which involves two imaginary-time intervals (of length $\beta/2$) as shown in Fig. 1(a) \cite{Schwinger,Keldysh,UmezawaAFT}. It will be argued how the external sources on these intervals univocally define states of the CFT, which can be described as pure bra/kets in the TFD framework
so that the CFT path integral describes a in/out scattering process. Finally, we will discuss   the gravitational dual to the thermal SK path in the large $N$ approximation, and present a new solution involving a two-sided black holes that solves the boundary problem and dominates over Thermal  AdS at high temperatures. We will also highlight the relation between the Lorentzian part of the geometry and the TFD-extended evolution operator in the dual field theory.
 
The SvR holographic prescription can be summarized in the following formula 
\begin{equation}\label{SvRPath}
Z_{CFT} \left[\phi({\cal C})\right]
=
Z_{grav}\left[\Phi|_{\partial}=\phi\left({\cal C}\right)\right] \;\;\;
\end{equation}
where the lhs is the generating function for correlation functions of CFT operators $\cal O$ with the sources $\phi $ having support on any continuous  path ${\cal C}$ in the complex $t$-plane. The rhs is the partition function for the bulk field $\Phi$, dual to $\cal O$, on an aAdS spacetime with asymptotic boundary conditions $\phi $.  

This is a remarkable path integral expression that captures all possible spacetimes combining regions of both signatures
for a specific contour choice ${\cal C}$ 
\cite{SvRC,SvRL}.
In particular, it applies to the purely Euclidean set up \cite{GKP,W},  and e.g. the current closed-contour case. In the Schwinger-Keldysh context the path $\cal C$ is closed and the lhs of \eqref{SvRPath} is expressed as follows
\begin{equation}\label{ZCFT}
Z_{CFT} = \text{Tr} \, \;U \qquad\qquad U \equiv {\cal P} \,e^{-i\int_{{\cal C}_{}} d\theta\; (H + {\cal O} \,\phi(\theta))}
\end{equation}
where $U$ is the evolution operator for a (CFT) Hamiltonian deformed with the source $\phi$.

Here we consider the path in Fig. \ref{Fig:Camino}(a): it consists in four intervals and the evolution operator factorizes as  $U = U_L U_{F} U_R U_{I}$, where $U_{L/R}$ are ordinary real time evolution operators, and $U_{I,F}$, on imaginary time intervals, are naturally associated to the states of the system at different times. The path ordering ${\cal P}$ defined in Fig. \ref{Fig:Camino}(a), corresponds to the arrowed lines depicted in Fig. \ref{Fig:Camino}(b), on the boundary of the spacetime.

The simplest check for this interpretation is by taking $T_{I,F} \to 0$ and vanishing sources, then one recovers the conventional thermal density matrix 
\begin{equation}\label{UU}
\rho_0 \equiv \lim_{\phi_{I,F}\to0} \;U_F \, U_{I}= e^{-\beta H}\;.
\end{equation}
In the next section we will show that via the TFD formalism we can associate the operators
\begin{equation}\label{Upm}
 U_{I,F} \equiv {\cal P} \,e^{-i\int_{I,F} d\theta \; (H + {\cal O} \,\phi(\theta))}
\end{equation}
to the in/out states of a scattering process in a Hilbert space duplicated according to the TFD rules. Notice that $\theta$ is pure imaginary in this sections. The matrix elements of $U_{I,F}$ represent amplitudes so as pure states do, and its explicit connection with TFD \textit{kets} is described below.

\begin{figure}[t]\centering
\begin{subfigure}{0.49\textwidth}\centering
\includegraphics[width=.9\linewidth] {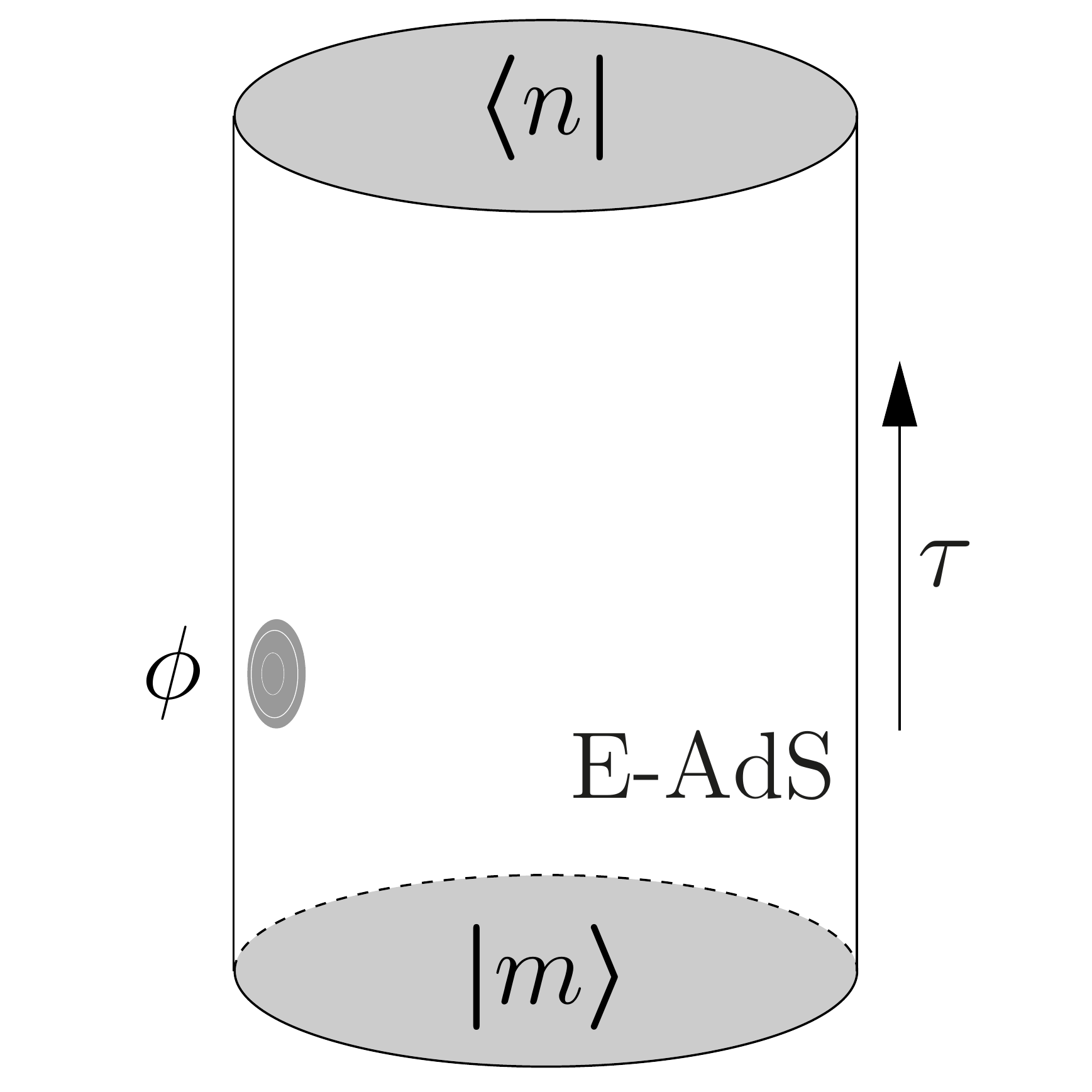}
\caption{}
\end{subfigure}
\begin{subfigure}{0.49\textwidth}\centering
\includegraphics[width=.9\linewidth] {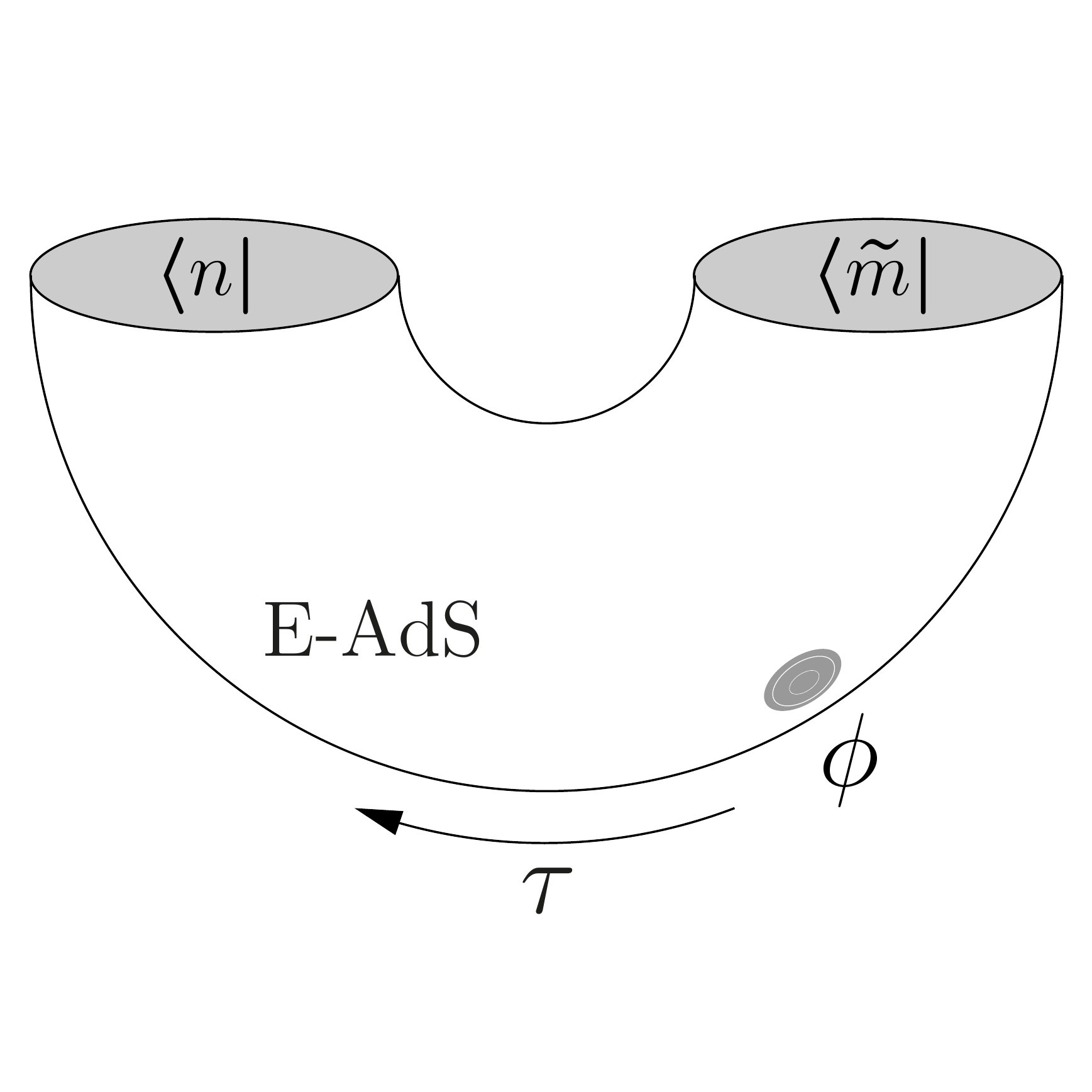}
\caption{}
\end{subfigure}
\caption{(a) A finite Euclidean time evolution 
depicted in terms of matrix elements $\langle n |U_\phi| m \rangle$. 
(b) The same geometry can be also understood as the component $\left(\langle n|\otimes \langle \tilde m | \right)\;|\Psi_{\phi} \rangle\!\rangle$ of a ket $|\Psi_\phi \rangle\!\rangle$ defined in the TFD Hilbert space $\cal H \otimes\tilde H$.}
\label{Fig:States}
\end{figure}

\subsection{ The TFD formulation and the in/out scenario }

The Thermo-Field Dynamics (TFD) formalism \cite{Takashi-TFD} is entirely equivalent for the computation of time ordered correlation functions at finite temperature to the conventional Schwinger-Keldysh (SK) method with $\sigma=\beta/2$ \cite{TFDUme}. 
Nevertheless, an expected feature of the TFD formulation is that deviations from the thermal vacuum state constructed by inserting sources in the imaginary-time intervals, immediately get interpreted as in/out excitations of a scattering set up in a thermal bath.

Let $\langle n |U_{\phi}| m \rangle $ denote the matrix elements of the evolution operators in an Euclidean 
piece, say $I$ of eq. (\ref{Upm}), where $n,m$ belong to a complete basis of the Hilbert space ${\cal H}$ of the field theory (CFT). 
In the TFD formalism, one constructs a second copy of the system by taking the CPT conjugate\footnote{The precise rules to construct the tilde states/operators can be found in \cite{Takashi-TFD}}, namely $\widetilde{{\cal H}}$, so that the total new system consist of the original CFT and its TFD copy living on disconnected asymptotic boundaries of the gravity dual, whose total states space is ${\cal H}\otimes\widetilde{{\cal H}}$.
Within this context, the TFD in-states $|\Psi_{\phi} \rangle\!\rangle$ are defined as
\be\label{defTFDstate}
\left(\langle n|\otimes \langle \tilde m | \right)\;|\Psi_{\phi} \rangle\!\rangle \equiv \langle n |U_{\phi}| m \rangle
\ee
where $| n\rangle, |\tilde m \rangle $ are orthonormal basis of ${\cal H}$ and $\widetilde{{\cal H}}$ respectively.

Using this definition, the dual, $\langle\!\langle\Psi_{\phi}|$,  of an initial (excited) state constructed with the source $\phi(\tau)$ in the lower half of $S^1$ ($ \tau \in (-\pi, 0)$) is related to the adjoint of its matricial form: $(U_{\phi})^\dagger$, and this is obtained by defining 
on the upper half  ($\tau \in(0, \pi)$) of the circle  \cite{Jackiw}:  $\left(\phi(\tau)\right)^{*} \equiv \phi(-\tau)$. This relation has been argued in the holographic context in \cite{us}.

The solution to \eqref{defTFDstate} is
\be
\label{psi-rho-id}
|\Psi_\phi \rangle\!\rangle = (U_\phi\otimes \mathbb I)|1\rangle\!\rangle\;  =U_\phi\;|1\rangle\!\rangle\,, 
\ee
where the unit state $|1\rangle\!\rangle$ is defined as \cite{Santana}
\be\label{unit}
|1\rangle\!\rangle \equiv\sum_n \,|n\rangle \otimes |\tilde{n} \rangle\;.
\ee
This can be verified by noticing that $\langle \tilde{m}| (U_\phi\otimes \mathbb I)|1\rangle\!\rangle = U_\phi\langle \tilde{m}  |1\rangle\!\rangle = U_\phi |m\rangle $. 
Expression \eqref{defTFDstate} is schematically represented in Fig. \ref{Fig:States}: $U_\phi$ is depicted on the left as an evolution operator  on a single Hilbert space , and the corresponding TFD-ket  $|\Psi_\phi\rangle\!\rangle$ is illustrated on the right with its two ends now representing the d.o.f. of the TFD double intersected at some spacelike surface of fixed time $t$. 

From $|\Psi_\phi\rangle\!\rangle$ one can define an Hermitian (reduced) density matrix
\be
\label{densitystate}
\rho_\phi \equiv \text{Tr}_{\widetilde{{\cal H}}}\;\,|\Psi_\phi \rangle\!\rangle \langle\!\langle \Psi_\phi |= Tr_{\widetilde{{\cal H}}}\,\,\,U_\phi \, |1\rangle\!\rangle \langle\!\langle  1|\, U_\phi^\dagger = \,U_\phi\, \,U_\phi^\dagger\,,
\ee
where we have used
\be
\text{Tr}_{\widetilde{{\cal H}}}\,\,\, |1\rangle\!\rangle \langle\!\langle  1|\,= \sum_n \,\, |n\rangle \langle n| = \mathbb{I}_{{\cal H}}\,.
\ee
Hermiticity of $\rho_\phi$ follows immediately
\be
\rho_\phi^\dagger \equiv \left( U^{}_\phi \, \, U_\phi^\dagger \right)^\dagger =  \left( U_\phi^\dagger \right)^\dagger \,\,U_\phi^\dagger = \rho^{}_\phi
\ee
where in the last step, we have transposed and then commuted.
These expressions explicitly show the connection between the pure state \eqref{psi-rho-id}  in the TFD setup and the mixed matrix density \eqref{densitystate}   in a single Hilbert space, both univocally determined by the evolution operator $U_\phi$ through  $\beta/2$ Euclidean time.

Working in the large $N$ limit, we have shown in \cite{us}  that non-trivial Euclidean sources $\phi$ on open contours  lift the system from the vacuum to coherent excited states,  with eigenvalues essentially given by the Fourier modes of $\phi$.  In the present context, we can carry a similar analysis to conclude that holographic excitations, built by imposing sources in the Euclidean boundaries as in \eqref{Upm} and \eqref{psi-rho-id}, correspond to \emph{thermal} coherent states \cite{Thermal-Coherent} in the bulk Fock space at large $N$ \cite{EssayBoots}.
However, we postpone the analysis of Euclidean sources  to an upcoming paper \cite{tocome}.

\begin{figure}[t]\centering
\begin{subfigure}{0.49\textwidth}\centering
\includegraphics[width=.9\linewidth] {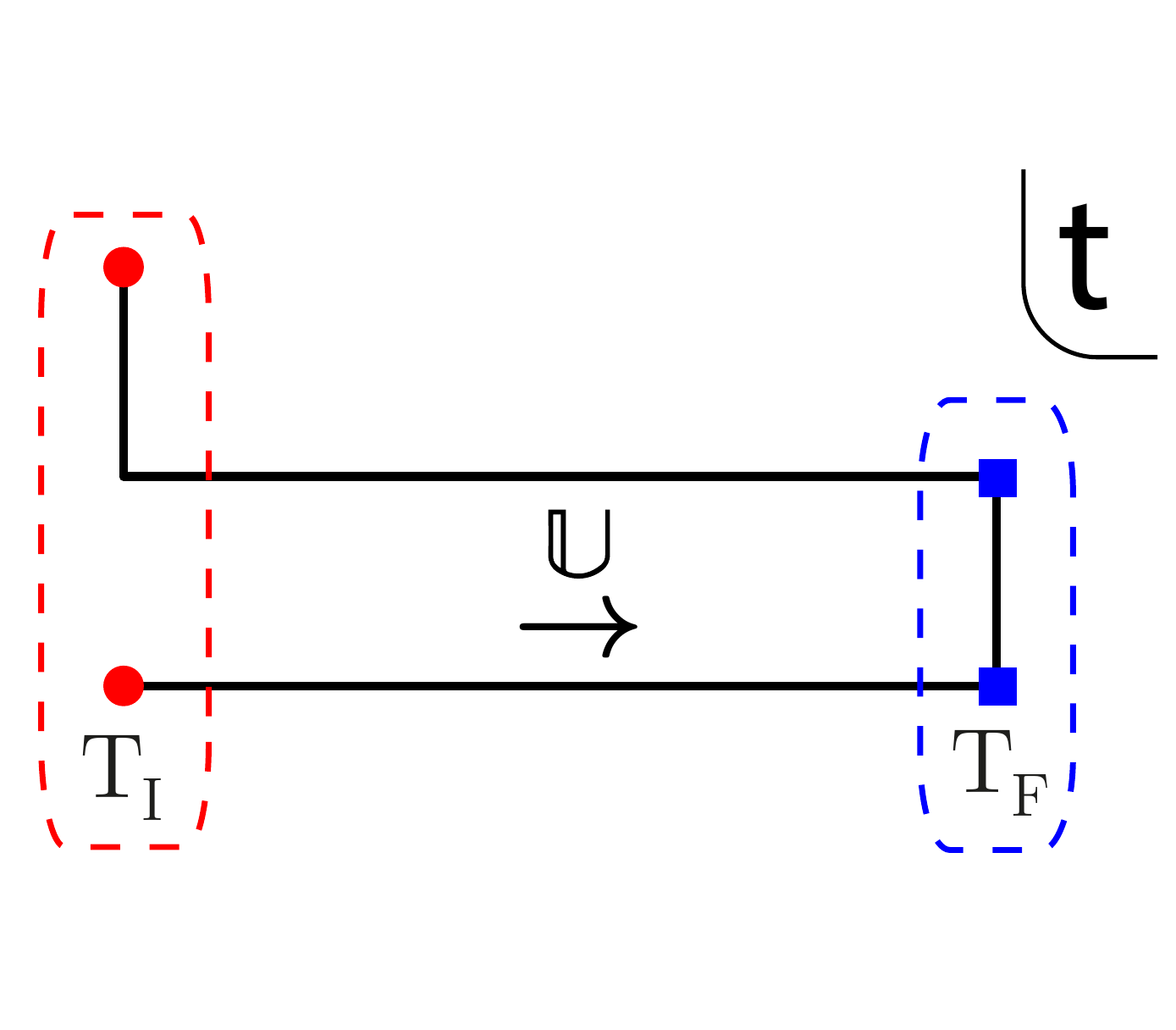}
\caption{}
\end{subfigure}
\begin{subfigure}{0.49\textwidth}\centering
\includegraphics[width=.9\linewidth] {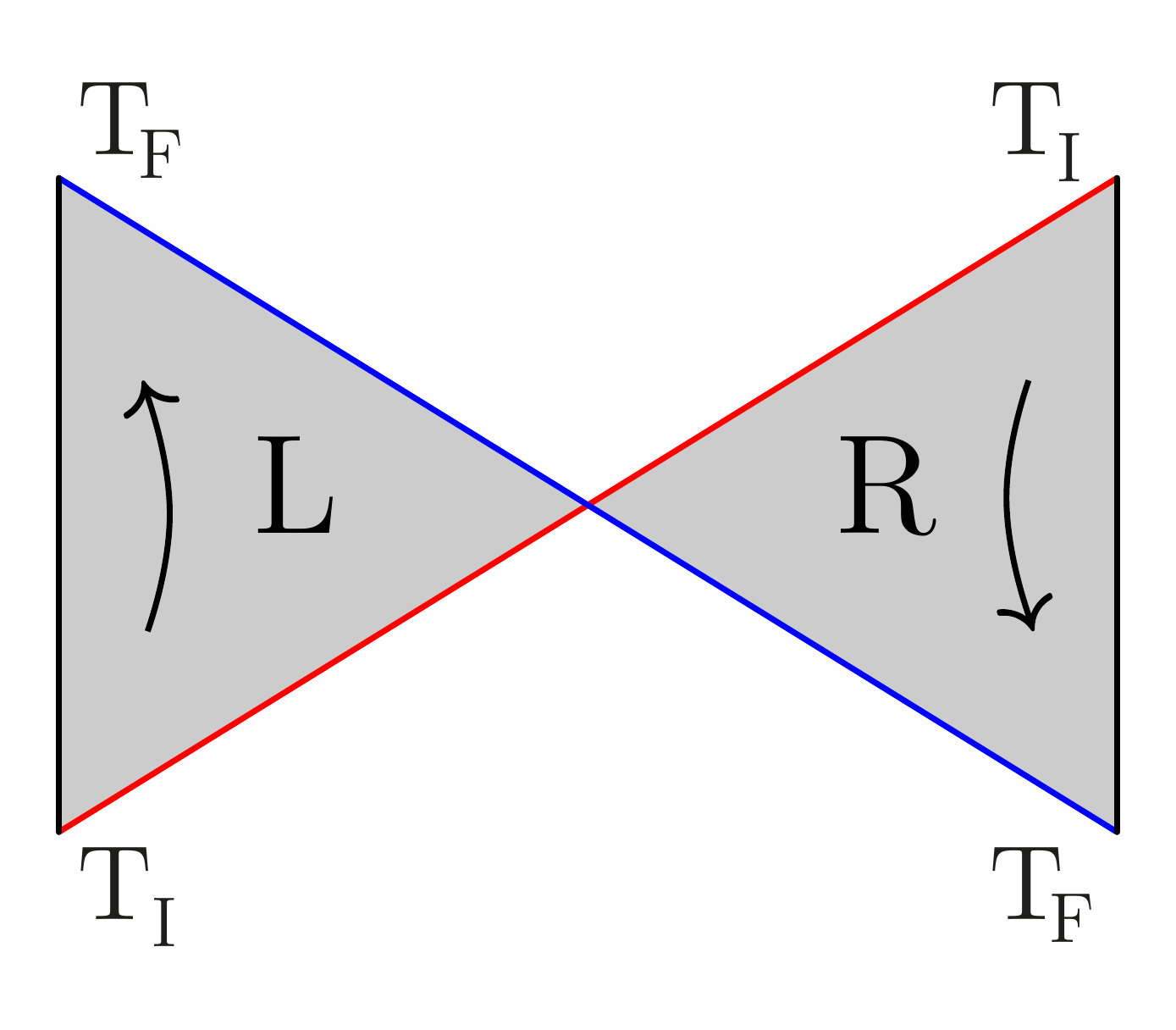}
\caption{}
\end{subfigure}
\caption{(a) Schwinger-Keldysh path in the complex $t$-plane. The vertical lines set up the initial and final states, while the horizontal lines represent the real time evolution of the TFD system.
(b) Two-sided exterior regions of the AdS-BH showing the surfaces on which we insert the in/out-wave functions. The arrows depict the boost like character of the TFD evolution.}
\label{Fig:Boots}
\end{figure}

We would like to conclude this section with two important observations. First,  by virtue of (\ref{psi-rho-id}), as $\phi \to 0$ the state  reduces to the TFD vacuum:
\be
\label{vacuumTFDstate}
\lim_{\phi^{}\to0} |\Psi_{\phi} \rangle\! \rangle = e^{- \frac \beta 2\, H } \,| 1 \rangle\! \rangle \equiv |\Psi_0 \rangle\! \rangle \;.
\ee 
As shown in \cite{eternal}, the high-temperature gravity dual of \eqref{vacuumTFDstate} is the Hartle-Hawking wave functional which in the semi-classical approximation is given by half of the Euclidean black hole solution.
Secondly,  the on-shell action evaluated on the Lorentzian part of the solution (see Fig. \ref{Fig:Boots}) nicely represents the CFT evolution operator that connects the initial/final states, namely 
\be\label{UTFD}
\mathbb{U} \equiv {\cal P}\, U_L \otimes U_R = e^{-i \, \Delta t \,\left( H \otimes\, \mathbb I - \mathbb I\, \otimes  H \right)} = e^{ -i\, \Delta t \,
\,\left(H - \tilde{H}\right)}\;,\ee where $\Delta t \equiv T_F -T_I \geq 0$ 
corresponds to the standard boost-like time in the gravity dual.
The operator (\ref{UTFD}) is obtained by considering only the path ordering of the two real time components of ${\cal C}$. It preserves the state $|\Psi_0 \rangle\! \rangle$.

Summarizing, the TFD framework re-interprets the CFT partition function as an In-Out process in a duplicated space. Combining expressions \eqref{vacuumTFDstate} and \eqref{UTFD} allow to express the l.h.s of \eqref{ZCFT} as
\begin{equation}
Z_{CFT}=\langle\!\langle \Psi_0 |\;\mathbb{U}[\phi_L,\phi_R]\;|\Psi_0 \rangle\! \rangle\,,
\end{equation}
where Lorentzian sources $\phi_{L/R}$ shall be considered for computational purposes. These remarks will be studied more in-depth and generalized to include excited states in 
\cite{tocome}.

\subsection{The Schwinger-Keldysh boundary problem in the large $N$ limit}

The CFT path integral over fields on the Schwinger-Keldysh closed path described above, see Fig. \ref{Fig:Camino}(a), corresponds to consider all non-trivial gravitational solutions with such contour as asymptotic boundary.

Noticeably, this problem has two classical solutions such as in the purely Euclidean closed path: the first one is the pure AdS torus whose global time coordinate must be substituted by the complex parameter of the path $\cal C$ (see Fig. \ref{Fig:ThermalAdS} below); and the second one is the solution presented in this work. It involves the exterior of the maximally extended AdS black hole in real time properly glued with Euclidean black hole pieces as depicted in Fig.\ref{Fig:Camino}(b).

On these gravitational backgrounds, one may also consider, for example, a non-back reacting scalar field with aAdS boundary conditions given by the SK path in the CFT. In the large $N$ semi-classical limit, the holographic recipe simplifies by using the saddle point approximation on the right hand side of \eqref{SvRPath}. This can be expressed as
\be\label{SvRpath + large N}
Z_{CFT}[\phi({\cal C})] \approx e^{-S_{AdS}[\phi({\cal C})]} + e^{-S_{BH}[\phi({\cal C})]}
\ee
where each term is the exponential of the on-shell action valued on each of both aAdS solutions, depending only on the boundary data $\phi$ on $S^d \times {\cal C}$. They are topologically distinct solutions, and the second term based in the new geometry, is the main object in our study. This situation resembles the conventional description of the Hawking-Page phase transition but now with a real time interval $t\in\left[ T_I , T_F\right]$ inserted between the two Euclidean halves (see Fig. \ref{Fig:Camino}). 
The bulk Euclidean actions are nothing but the free energies of each solution, determining the preferred background at a given $\beta$, while the Lorentzian terms  provide the $n$-point correlation functions for boundary operators.  The high/low temperature regimes are dominated by the BH/Thermal solution.

In the forthcoming sections we study the dual geometries to Fig.\ref{Fig:Camino}(a) via 2-pt correlators for   a non-backreacting scalar field. The causal properties of these correlators are completely determined by the path ordering shown in Fig. \ref{Fig:Camino}(a). Therefore, the structure of the lhs of eq. \eqref{ZCFT} is known providing expressions to directly compare the bulk results we are going to compute. Differentiation wrt the Lorentzian sources $\phi_{l}$, $l=\{L,R\}$, give the Schwinger-Keldysh propagator,
\begin{equation}\label{PropMatrix-CFT}
-i\,\frac{\delta^2\ln Z_{CFT}}
{\delta\phi_l\,\delta\phi_{l'}}
\equiv \int d^d\!\!k \; e^{-i k (x-y) } \langle\!\langle\Psi_0| \left(\begin{array}{cc} {\cal O}_L (k) {\cal O}_L (k) & {\cal O}_L (k) {\cal O}_R (k) \\ {\cal O}_R (k) {\cal O}_L (k) & {\cal O}_R (k) {\cal O}_R (k) \end{array}\right) |\Psi_0\rangle\!\rangle \ ,
\end{equation}
where the matrix elements are
\begin{align}\label{SKprop}
\langle\!\langle\Psi_0|{\cal O}_L (k) {\cal O}_L (k )|\Psi_0\rangle\!\rangle &=\frac{e^{\beta\omega}} {e^{\beta\omega}-1} G_R(k) -\frac{1} {e^{\beta\omega}-1} G_A(k) \,,\\
\langle\!\langle\Psi_0|{\cal O}_R (k) {\cal O}_L (k)|\Psi_0\rangle\!\rangle &= 2 \frac{ e^{\beta\omega/2}} {e^{\beta\omega}-1}\,\left(G_A(k)-G_R(k)\right)=\langle\!\langle\Psi_0|{\cal O}_L (k) {\cal O}_R (k)|\Psi_0\rangle\!\rangle\,, \label{offdiag}\\
\langle\!\langle\Psi_0|{\cal O}_R (k) {\cal O}_R (k)|\Psi_0\rangle\!\rangle &= \frac{1}{e^{\beta\omega}-1} G_R(k) -\frac{e^{\beta\omega}} {e^{\beta\omega}-1} G_A(k)\,.\nn
\end{align}
In the expressions above, $G_{R/A}(k)$ stand for the Fourier transform of the retarded and advanced propagators of the theory, analytic in the upper and lower half of the complex $\omega\equiv k^0$ plane. The symmetric property of the matrix is a consequence of the two Euclidean pieces having equal $\beta/2$ length. The general matrix was presented in \cite{herzog}, where this particular path was already seen to appear naturally in gravity.

\section{Building up the gravity solution}
\label{DualGeom}

In this section we describe in detail the construction of the geometry, depicted in Fig. \ref{Fig:Camino}(b), dual to the SK path shown in Fig. \ref{Fig:Camino}(a). We will define the coordinates we will work with and take care of the gluing conditions between the different signature regions. We will work in $2+1$ dimensions for the ease of calculation, but the whole construction follows straightforwardly to higher dimension examples.

\begin{figure}[t]\centering
\begin{subfigure}{0.49\textwidth}\centering
\includegraphics[width=.9\linewidth] {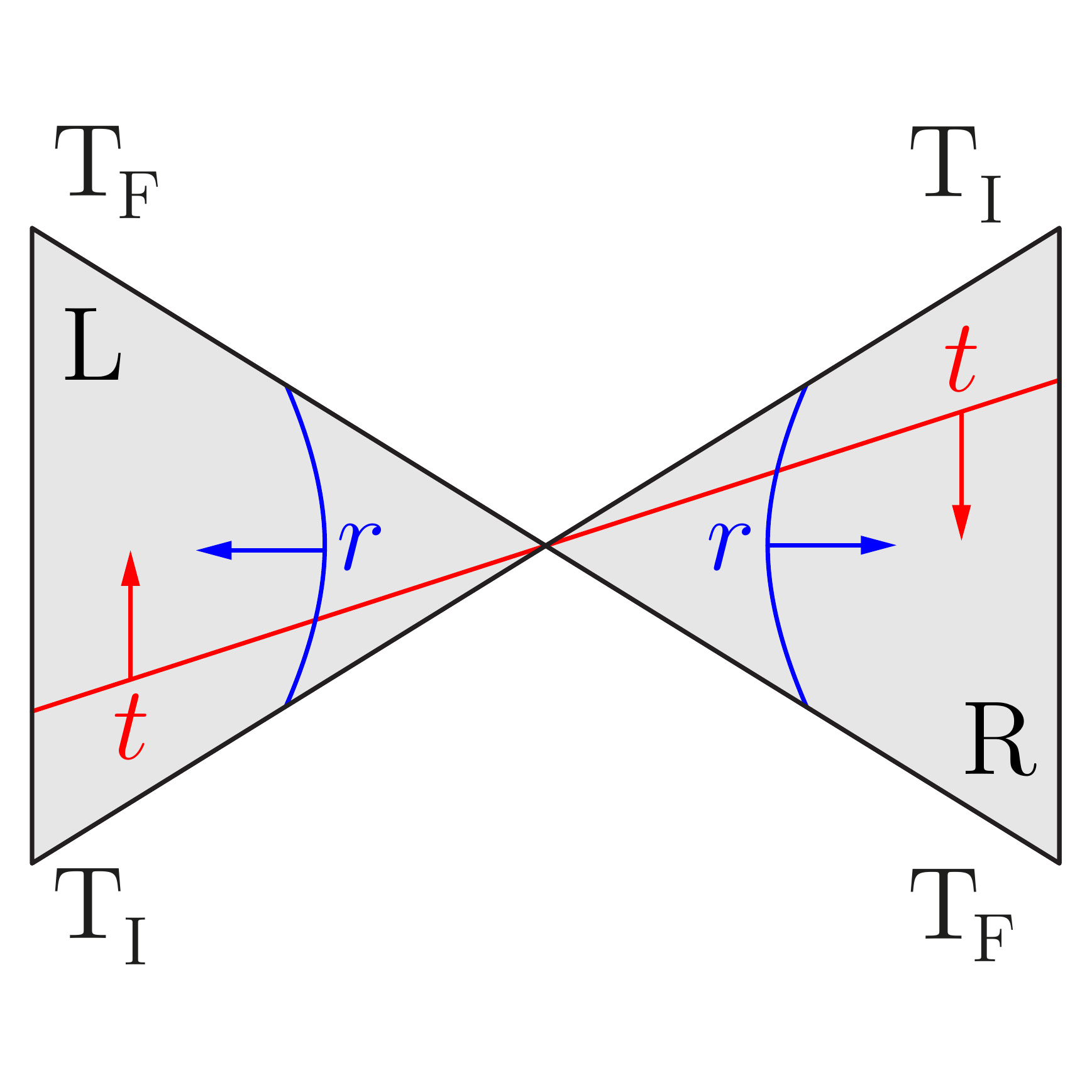}
\caption{ }
\end{subfigure}
\begin{subfigure}{0.49\textwidth}\centering
\includegraphics[width=.9\linewidth] {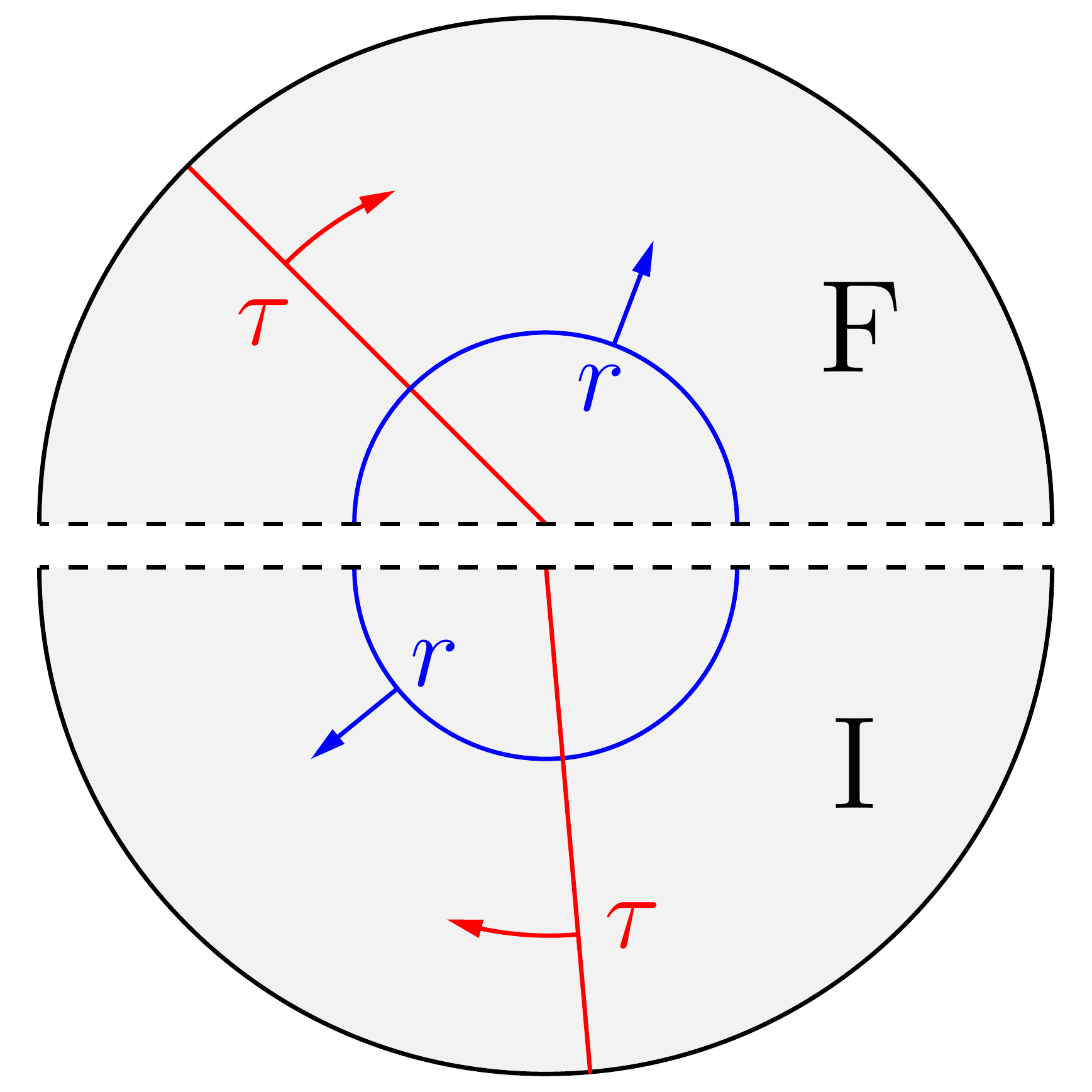}
\caption{}
\end{subfigure}
\caption{(a) Static patches of the AdSBH with constant $t,r$ surfaces depicted. Time runs upward in the left wedge (L) and downward in the right wedge (R). The angular variable $\varphi$ in \eqref{metric-l1} has been suppressed. 
(b) Euclidean AdSBH: time becomes an angular variable $\tau\sim\tau+2\pi$. The two pieces are identical and their temporal extension is $\beta/2$. }
\label{Fig:Guille}
\end{figure}

The geometry is built from the Lorentzian AdS-BH exteriors L and R and the Euclidean BH manifold halved in two pieces as shown in Fig.\ref{Fig:Guille}. The two Euclidean pieces are glued to the constant $t$-hypersurfaces (red lines in Fig. \ref{Fig:Guille}(a)) located at $t=T_\pm$.
The standard metrics for the BTZ black hole are ($R_{AdS}\equiv1$) \cite{BTZ}
\begin{equation}\label{metric-l1}
ds^2=-\left(  r^2-r_S^2\right)d t^2+\frac{d  r^2}{  r^2-r_S^2}+  r^2 d \varphi^2
\qquad \text{and}\qquad
ds^2=\left(  r^2-r_S^2\right)d \tau^2+\frac{d  r^2}{  r^2-r_S^2}+  r^2 d \varphi^2\,,
\end{equation}
in Lorentzian and Euclidean signature respectively. In these metrics, $  t\in\mathbb{R}$, $ \varphi\sim \varphi +2\pi$ and $ \tau\sim  \tau +\beta$ with $\beta=T^{-1}=2\pi/r_S$.  Rescaling the coordinates as
\begin{equation}\label{rescaling}
  r \to r_S\,r\;,\qquad   t \to\frac{t}{r_S}\;,\qquad \tau\to\frac{\tau}{r_S}\;,\qquad \varphi\to \frac{\varphi}{r_S}\;,
\end{equation}
turn the metrics \eqref{metric-l1} into
\begin{equation}\label{metric-l2}
ds^2=-(r^2-1)dt^2+\frac{dr^2}{(r^2-1)}+r^2 d\varphi^2 \qquad\qquad ds^2=(r^2-1)d\tau^2+\frac{dr^2}{(r^2-1)}+r^2 d\varphi^2
\end{equation}
with $\tau\sim \tau +2\pi$ and 
the BH temperature absorbed in the angular periodicity $\varphi\sim\varphi+2\pi r_S$. We will work with metrics \eqref{metric-l2} for our geometry throughout this paper.

\subsection{Continuity conditions}

We  now show that our geometry meets the appropriate ${\cal C}^1$ gluing conditions, this is, continuity of the metric and extrinsic curvature across constant time surfaces. 
The metrics on any constant Lorentzian and Euclidean time slices coincide and can therefore be continuously glued. The staticity of the spacetime guarantees that this can be done at any value\footnote{By writing  metrics \eqref{metric-l2} in Kruskal coordinates one can check continuity at $u=0$ and $v=0$ surfaces, i.e. in the $t\to\infty$ limit.} of $t$ and $\tau$. The continuity of the conjugate momentum of the metric is equivalent to demand continuity of the extrinsic curvatures $K_{\mu\nu}$ across the gluing surface \cite{SvRL}. Explicitly,
\begin{equation}\label{extcurv}
K_{\mu\nu}\equiv\frac 12 {\cal L}_n P_{\mu \nu} = \frac 12 n^\alpha \partial_\alpha \left(g_{\mu\nu}-n^2 n_{\mu}n_{\nu}\right) + \partial_\mu \left(n^\alpha\right)  \left(g_{\alpha\nu}-n^2 n_{\alpha}n_{\nu}\right)
\end{equation}
where $P_{\mu\nu}=g_{\mu\nu}-n^2 n_{\mu}n_{\nu}$ is the first fundamental form of a hypersurface with normal $n^\mu$. In the present case, the unitary time-like vector is $n^\mu\equiv \delta_0^\mu\,(r^2-1)^{-1/2}$. The first term in \eqref{extcurv} vanishes due to staticity while the second gives
\begin{align} 
K_{\mu\nu}&=\partial_\mu \left(n^\alpha\right)  \left(g_{\alpha\nu}-n^2 n_{\alpha}n_{\nu}\right)
= \delta_\mu^r \left( \frac{-r}{(r^2-1)^{\frac 32}}\right)\left(g_{0\nu}-n^2n_0n_\nu \right)=\delta_\mu^r\delta_\nu^0\left( \frac{-r}{(r^2-1)^{\frac 32}}\right)P_{00}=0
\end{align}
which follows from $P_{00}=0$. The expression above shows the ${\cal C}^1$ continuity of the metric across the $\Sigma_t$ and $\Sigma_\tau$ gluing surfaces. The point $r=1$ shows no special pathology, as $\lim_{r\to1}K_{\mu\nu}=0$.

We stress that the L and R sections, depicted in Fig. \ref{Fig:Guille}, are connected through the wormhole, located at $r=1$, as in the standard Kruskal extension. This gluing is a natural assumption to avoid boundary conditions at $r=1$ when solving for the bulk field. The final outcome of the construction is the geometry depicted in Fig. \ref{Fig:Camino}(b). 

Regarding the parametrization on each region, we define $t$ on L and R as the real part of their respective position in the complex $t$ plane of Fig. \ref{Fig:Camino}(a), i.e. $t\in\left[ T_I,T_F\right]$ both in L and R running from left to right. We will take $-T_I=T_F=T/2$ to simplify notation in the explicit computation. The Euclidean regions are parametrized $\tau\in[0,\pi]$ in $F$ and $\tau\in[-\pi,0]$ in $I$.

We now discuss the boundary conditions on the fields defined on this geometry. These follow directly from the saddle point approximation on the rhs of \eqref{SK+CFT} and can be understood as a ${\cal C}^1$ gluing of the fields, i.e. continuous field and conjugated momentum, through the (bulk) hyper-surfaces joining the pieces of the SK path. The explicit signs in the gluing conditions, specifically on the conjugated momentum, depend on the time parametrization on each region. Our choice is shown in Fig. \ref{Fig:Guille}(a) and leads to
\begin{eqnarray}\label{bc}
\Phi_L=\Phi_I\,,  & \quad
-i\partial_t\Phi_L=\partial_\tau\Phi_I\,, 
& \qquad\text{on $t=T_I$, $\tau=0$}\nn \\
\Phi_L=\Phi_F\,, & \quad
-i\partial_t\Phi_L=\partial_\tau\Phi_F\,, 
& \qquad\text{on $t=T_F$, $\tau=0$} \nn\\
\Phi_R=\Phi_I\,,& \quad
-i\partial_t\Phi_R=\partial_\tau\Phi_I\,, 
&\qquad\text{on $t=T_I$, $\tau=-\pi$} \nn\\
\Phi_R=\Phi_F\,,& \quad
-i\partial_t\Phi_R=\partial_\tau\Phi_F\,, 
& \qquad\text{on $t=T_F$, $\tau=\pi$}
\end{eqnarray}
We refer the reader to \cite{SvRL,us} for explicit computations in several examples. Notice that the time coordinate in R runs opposite to the path ordering.
This is in agreement with the TFD interpretation of the dofs in $R$ as being the CPT dual of those in region $L$. 

\section{Bulk Massive Scalar Field}
\label{BH}

In the present section we will find the classical bulk solutions for a real massive scalar field in  the two geometries dual to the SK path shown in Fig. \ref{Fig:Camino}(a) (cf. \eqref{SvRpath + large N}). The field will be subject to arbitrary boundary conditions on the Lorentzian asymptotic regions. This will provide us with the necessary data to reproduce the high and low temperature behavior of the propagator matrix \eqref{PropMatrix-CFT}.

\subsection{Two-sided black hole geometry }

\begin{figure}[t]\centering
\begin{subfigure}{0.49\textwidth}\centering
\includegraphics[width=.9\linewidth] {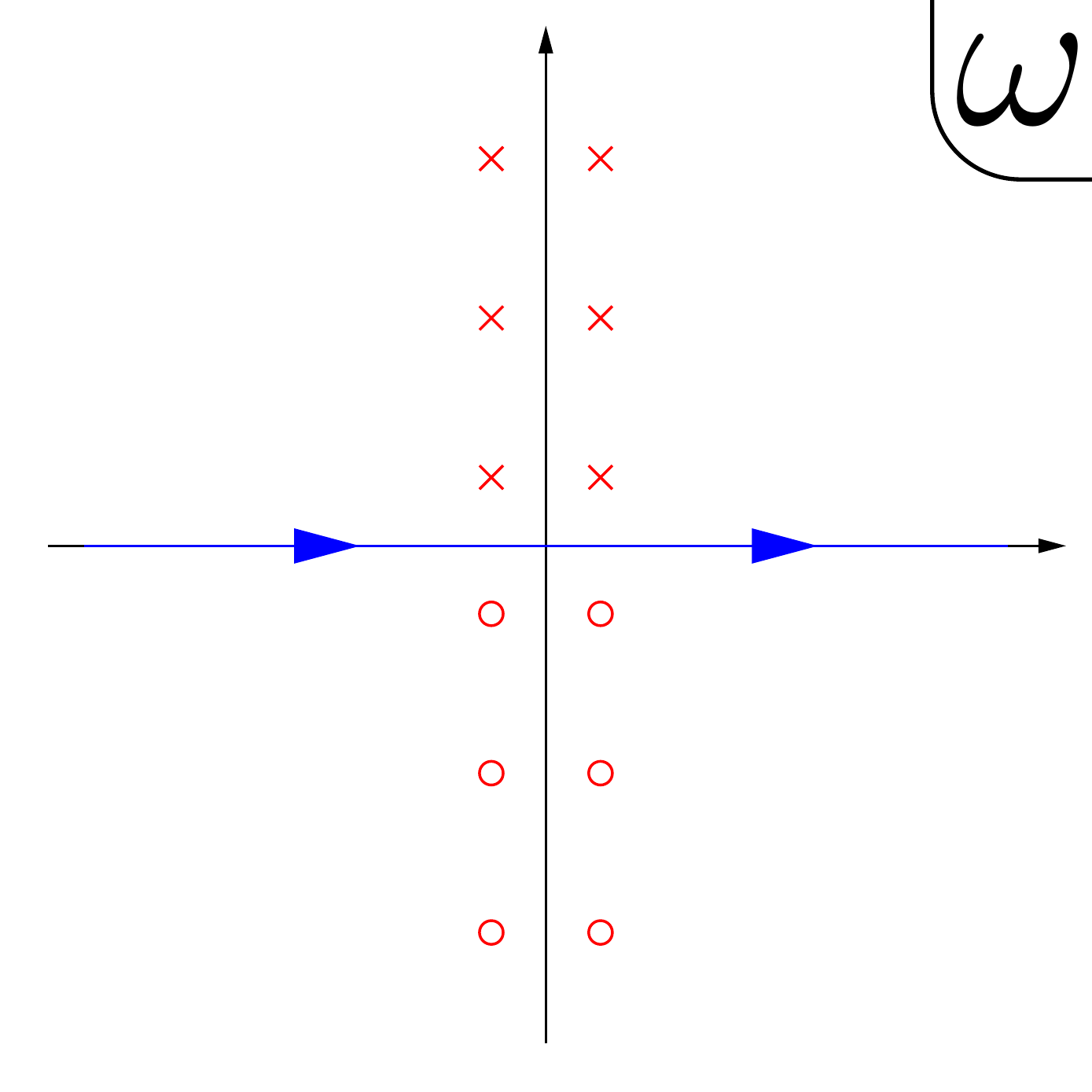}
\caption{}
\end{subfigure}
\begin{subfigure}{0.49\textwidth}\centering
\includegraphics[width=.9\linewidth] {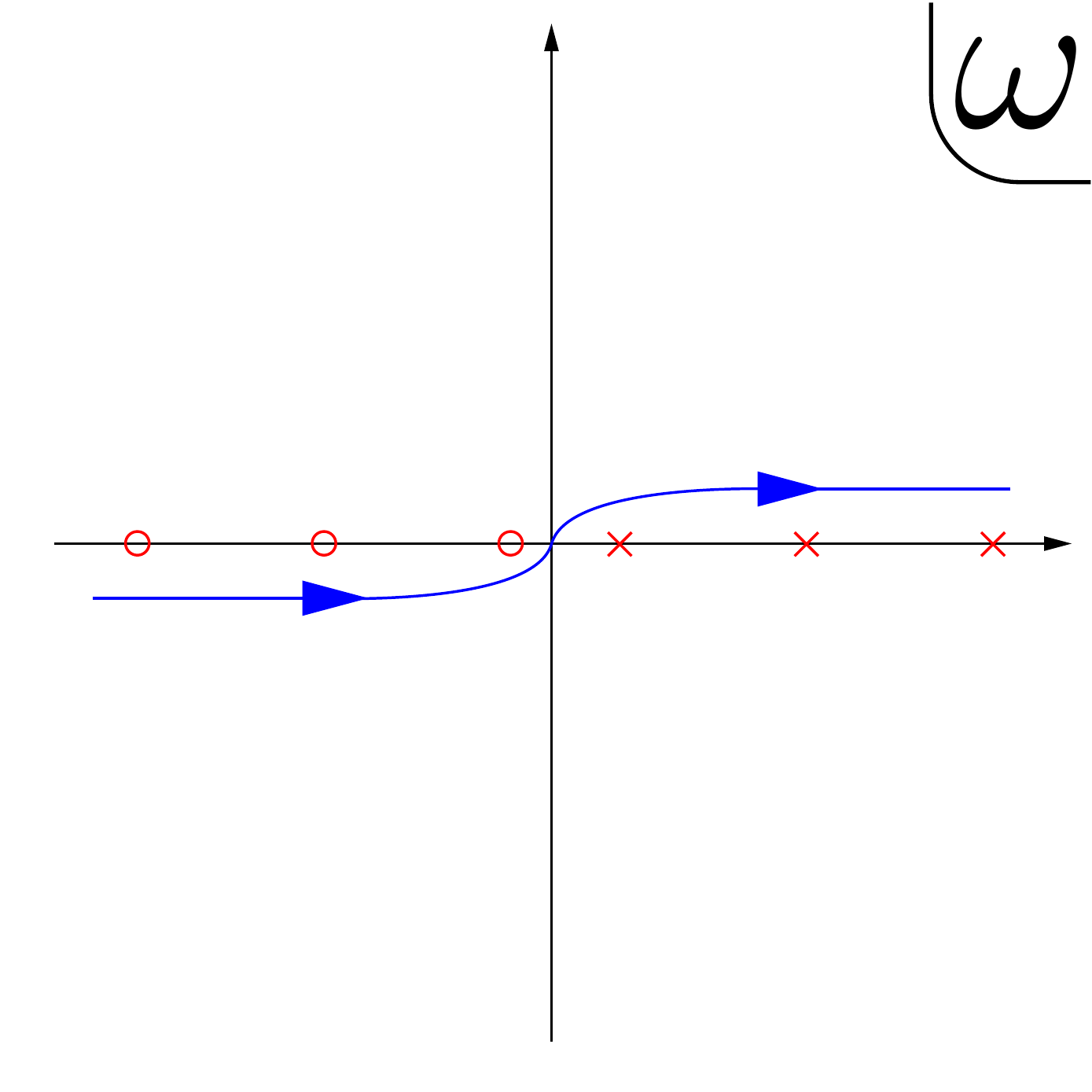}
\caption{}
\end{subfigure}
\caption{(a) The blue line denotes the $\omega$-integration contour in the Lorentzian solution \eqref{Lsol}. Crosses show the location of the poles of $f(\omega,l,r)$ while circles those of $f(-\omega,l,r)$. 
(b) The Feynman-like blue line represents the integration path in \eqref{ThermalLSol}. The red crosses and circles represent the poles of $s(\omega,l,r)$, which lie on the real axis.}
\label{Fig:Polos}
\end{figure}

We now proceed to build the solution for the scalar field $\Phi$ on the complete geometry with non-zero source $\phi_L$ on the L wedge. The solution with non-zero $\phi_R$ can be found following an analogous procedure, and the full solution is easily obtained due to the linearity of the problem. The action and EOM are given by
\begin{equation}
\label{scalarEOM}
S[\Phi]=-\frac 12 \int dt dr d\varphi \sqrt{|g|}\left(\partial_\mu\Phi\partial^\mu\Phi+m^2\Phi^2\right)\;,\qquad\qquad\left(\square-m^2\right)\Phi=0 \;,
\end{equation}
in the Lorentzian metric \eqref{metric-l2}. The field is subject to  $\Phi \sim r^{\Delta-2}\phi_L(t,\varphi)$ on region L, where $\Delta=1+\sqrt{1+m^2}$, and trivial sources everywhere else. Writing $\Phi = e^{-i \omega t + i l \varphi} f(\omega,l,r)$, where $r_S l \in\mathbb{Z}$, one obtains from \eqref{scalarEOM}
\begin{equation}\label{f}
f(\omega,l,r)\equiv {\cal N}_{\omega l\Delta} \; r^{-\Delta } \left(1-\frac{1}{r^2}\right)^{i\frac{\omega }{2}} \, _2F_1\left(\frac{\Delta }{2}+\frac{1}{2} i (\omega -l),\frac{\Delta }{2}+\frac{1}{2} i
   (\omega+l );i \omega +1;1-\frac{1}{r^2}\right)\,,
\end{equation}
\begin{equation}\nn
{\cal N}_{\omega l\Delta}\equiv \frac{\Gamma \left(\frac{\Delta }{2}+\frac{1}{2} i (\omega -l)\right) \Gamma \left(\frac{\Delta }{2}+\frac{1}{2} i (\omega +l)\right)}{\Gamma (\Delta -1) \Gamma (i \omega +1)}\,.
\end{equation}
The normalization factor ${\cal N}_{\omega l\Delta}$ fixes the asymptotic behavior of the solution to be\footnote{We remind the reader that the $\ln(r^2)$ term appears only for $\Delta\in\mathbb{N}$ and becomes relevant in KK compactifications. This will not be important in our discussion. We refer the interested reader to \cite{DHokerFriedman,Muck} and appendices in \cite{us2}.},
\begin{equation}\label{fexpansion}
f(\omega,l,r)\approx r^{\Delta-2}+\dots+\alpha(\omega,l,\Delta)r^{-\Delta}\left[\ln(r^2)+\beta(\omega,l,\Delta)+\dots\right]\,,\qquad r\to\infty
\end{equation}
\begin{equation}\label{alpha}
\alpha_{\omega l\Delta}\equiv (-1)^{\Delta -1} \frac{ \left(\frac{2-\Delta }{2}+\frac{i}{2}  (\omega - l)\right)_{\Delta -1} \left(\frac{2-\Delta }{2}+\frac{i}{2}  (\omega + l)\right)_{\Delta -1}}{(\Delta -2)! (\Delta -1)!}\,,
\end{equation}
\begin{equation}\label{beta}
\beta_{\omega l\Delta}\equiv-\psi \left(\frac{\Delta }{2}+\frac{i}{2}  (\omega -l)\right)-\psi \left(\frac{\Delta }{2}+\frac{i}{2}  (\omega +l)\right)\,,
\end{equation}
where $(x)_y$ stands for Pochhammer symbols and $\psi(x)$ for the Digamma function. It is important to note that the normalization factor introduces simple poles at $\omega=\pm l +i (2n+\Delta)$, with $n\in\mathbb{N}$. They are depicted by crosses in Fig. \ref{Fig:Polos}(a). Notice that $f(\pm\omega,l,r)$ results analytic in the lower/upper half plane respectively.
 
Contrary to the case of pure AdS, the BH geometry shows two linearly independent regular NN solutions:  $ e^{-i \omega t + il\varphi } f(\pm\omega,l,r)$. They can be chosen to behave as outgoing and ingoing waves at the horizon, respectively. A general solution on the L wedge is then built from,
\begin{equation}
\label{Lsol}
\Phi_L(r,t,\varphi)=\frac{1}{4\pi^2 r_S} \sum_{l} \int d\omega \int d t' d  \varphi' \; e^{-i\omega (t-t')+i l (\varphi- \varphi')} \phi_L(t',\varphi')\left[L^+_{\omega l} f(\omega,l,r) + L^-_{\omega l} f(-\omega,k,r)\right]\;,
\end{equation}
with $\omega\in\mathbb R$. To meet the asymptotic boundary condition we demand 
\be 
L^+_{\omega l} + L^-_{\omega l}=1\,.
\label{Lpm}
\ee
Although not mandatory, introducing $L^\pm_{\omega l}$ becomes handy for gluing the complete solution. Their difference parametrize the normalizable modes on the geometry. To gain physical intuition, the quotient $L^+_{\omega l}/L^-_{\omega l}$ can be interpreted as the relative weight of outgoing and infalling modes through the horizon in the NN solution.

Normalizable (N) modes are built from the combination $e^{-i \omega t +il\varphi } \left[f(\omega,l,r)-f(-\omega,l,r)\right]$ as is evident from \eqref{fexpansion}. These are the appropriate modes to expand the solution on regions R, I and F
\begin{align}
\Phi_R(r,t,\varphi)&= \frac{1}{4\pi^2 r_S} \sum_{l} \int d\omega \; e^{-i\omega t+i l \varphi} R_{\omega l} \left[ f(\omega,l,r) - f(-\omega,l,r)\right] \label{Rsol}\;,\\
\Phi_{F}(r,\tau,\varphi)&= \frac{1}{4\pi^2 r_S} \sum_{l} \int d\omega  \; e^{-\omega \tau+i l \varphi} \; \text{F}_{\omega l} \left[ f(\omega,l,r) - f(-\omega,k,r)\right]  \label{Isol}\;,\\
\Phi_I(r,\tau,\varphi)&= \frac{1}{4\pi^2 r_S} \sum_{l} \int d\omega \; e^{-\omega \tau+i l \varphi} \; \text{I}_{\omega l} \left[ f(\omega,l,r) - f(-\omega,k,r)\right] \label{IIsol}\;.
\end{align}
Since $\omega\in\mathbb R$,  \eqref{Isol} and \eqref{IIsol} seem to be ill defined at high frequencies, nevertheless, the gluing provides I$_{\omega l}$ and F$_{\omega l}$ which guarantee a regular solution.

We relegate the details of the gluing procedure to \cite{tocome}, where a general solution including non-zero Euclidean sources will be presented. Imposing \eqref{bc} on \eqref{Lsol}, \eqref{Rsol}-\eqref{IIsol}, one finds
\begin{align}\label{CoeffSol}
- L^-_{\omega l} \; \tilde\phi_L \; e^{-i \omega T/2}&=F_{\omega l}\;,
&~& 
F_{\omega l}\;e^{-\pi\omega}= \; R_{\omega l}\; e^{-i \omega T/2}\;, 
&~&
R_{\omega l}e^{i \omega T/2}  =I_{\omega l}\;e^{\pi\omega}\;,
&~& 
I_{\omega l}=L^+_{\omega l}e^{i \omega T/2}  \; \tilde\phi_L 
\end{align}
yielding via \eqref{Lpm}
\begin{equation}\label{AandB}
L^+_{\omega l}=\frac{-1}{e^{2\pi\omega}-1} \,, \qquad\qquad L^-_{\omega l}=\frac{e^{2\pi\omega}}{e^{2\pi\omega}-1}\;.
\end{equation}
The remaining coefficients can be obtained from \eqref{AandB} and \eqref{CoeffSol}.

~

\nin {\bf Comments}: 

\vspace{.2cm}

\nin {\sf 1. Uniqueness}: ~ ``{\it no source free solutions to the scalar field EOM exist}''.

Consider N modes in every region:
\begin{align}
\Phi_L(r,t,\varphi)&= \frac{1}{4\pi^2 r_S} \sum_{l} \int d\omega  e^{-i\omega t+i l \varphi} L_{\omega l} \left[ f(\omega,l,r) - f(-\omega,k,r)\right] \label{Lsol2}\;,
\end{align}
in L region, \eqref{Rsol}-\eqref{IIsol} in regions $R,I,F$   and impose \eqref{bc}. As a consequence one obtains
$$I_{\omega l}=e^{2\pi\omega}I_{\omega l}\;.$$
This condition on $I_{\omega l}$ can be understood as manifesting the $ -i\beta$ thermal periodicity of the geometry\footnote{Recall that the temperature was absorbed in the angular periodicity, turning $\beta=2\pi$  (cf. \eqref{metric-l2}). Thermal periodicities of $2\pi$ must be understood in units of $\beta$. \label{Footnote1}}. For arbitrary $\omega$ it implies $\text{I}_{\omega l}=0$, but for $\omega=im$, $m\in\mathbb{Z}$ the possibility of  arbitrary I$_{(im) l}$ arises. 
This set of solutions is nevertheless ruled out since the cancellation at infinity no longer works:  $ f(i m,l,r) = 0$ for $m>0$. We conclude that $\text{I}_{\omega l}=0$ is the only possible solution.

\vspace{.3cm}
\nin {\sf 2. Analyticity though the wormhole}: ~ ``{\it the Lorentzian solution results analytic in the two sided BH }''.

The solution above involved uncorrelated `Rindler' modes $\Phi_L$ and $\Phi_R$.  The gluing conditions \eqref{CoeffSol}, lead to
\begin{equation}\label{coeffsL}
R_{\omega l}=  \tilde\phi_L\; L^+_{\omega l}\; e^{\omega \pi }=-\;\tilde\phi_L\; L^-_{\omega l}\; e^{-\omega \pi }\,,
\end{equation}
relating the L and R coefficients. We would like to emphasize that the $e^{\pm\omega \pi}$ factors were not imposed a priori and arose as a consequence of regions I and II having a $\beta/2$ temporal extension$^{\ref{Footnote1}}$. It is well known that these relations are associated with global analytic modes \cite{Unruh}. This is one of the important outcomes of our construction: the $\beta/2$ extension of the Euclidean geometry imply that the fields in  the bulk end up being analytic. 
One could have carried out the computations above for other arbitrary Euclidean sections lengths, but the fields living in the Lorentzian regions would not have been analytic, thus invalidating the solution. Equal length Euclidean sections as naturally associated to BH geometries was noticed previously in \cite{herzog}. This property is particular to BHs, for the Thermal AdS setup we will see that no constraint arises, modes are analytic all over the geometry for any Euclidean temporal extension.

\vspace{.3cm}
\nin {\sf 3. Regularity }: ``{\it  Euclidean pieces \eqref{Isol} and \eqref{IIsol} become finite after the gluing}''

The final expressions for the bulk field in the Euclidean sections are
\begin{align}\nn
\Phi_F(r,\tau,\varphi)&= \frac{1}{4\pi^2 r_S} \sum_{l} \int d\omega  \; e^{-\omega \tau+i l \varphi} \; \left(-\tilde\phi_L\; e^{-i\omega T/2} \frac{e^{2\pi\omega}}{e^{2\pi\omega}-1}\right) \left[ f(\omega,l,r) - f(-\omega,k,r)\right] \;,\quad \tau\in(-\pi,0)\\
\Phi_I(r,\tau,\varphi)&= \frac{1}{4\pi^2 r_S} \sum_{l} \int d\omega \; e^{-\omega \tau+i l \varphi} \; \left(-\tilde\phi_L\; e^{+i\omega T/2} \frac{1}{e^{2\pi\omega}-1}\right) \left[ f(\omega,l,r) - f(-\omega,k,r)\right]\;,\quad \tau\in(0,\pi  )\nn\;.
\end{align}
The resulting coefficients regulate adequately the $\omega$ integrals, thus validating our procedure.

\begin{figure}[t]\centering
\begin{subfigure}{0.49\textwidth}\centering
\includegraphics[width=.9\linewidth] {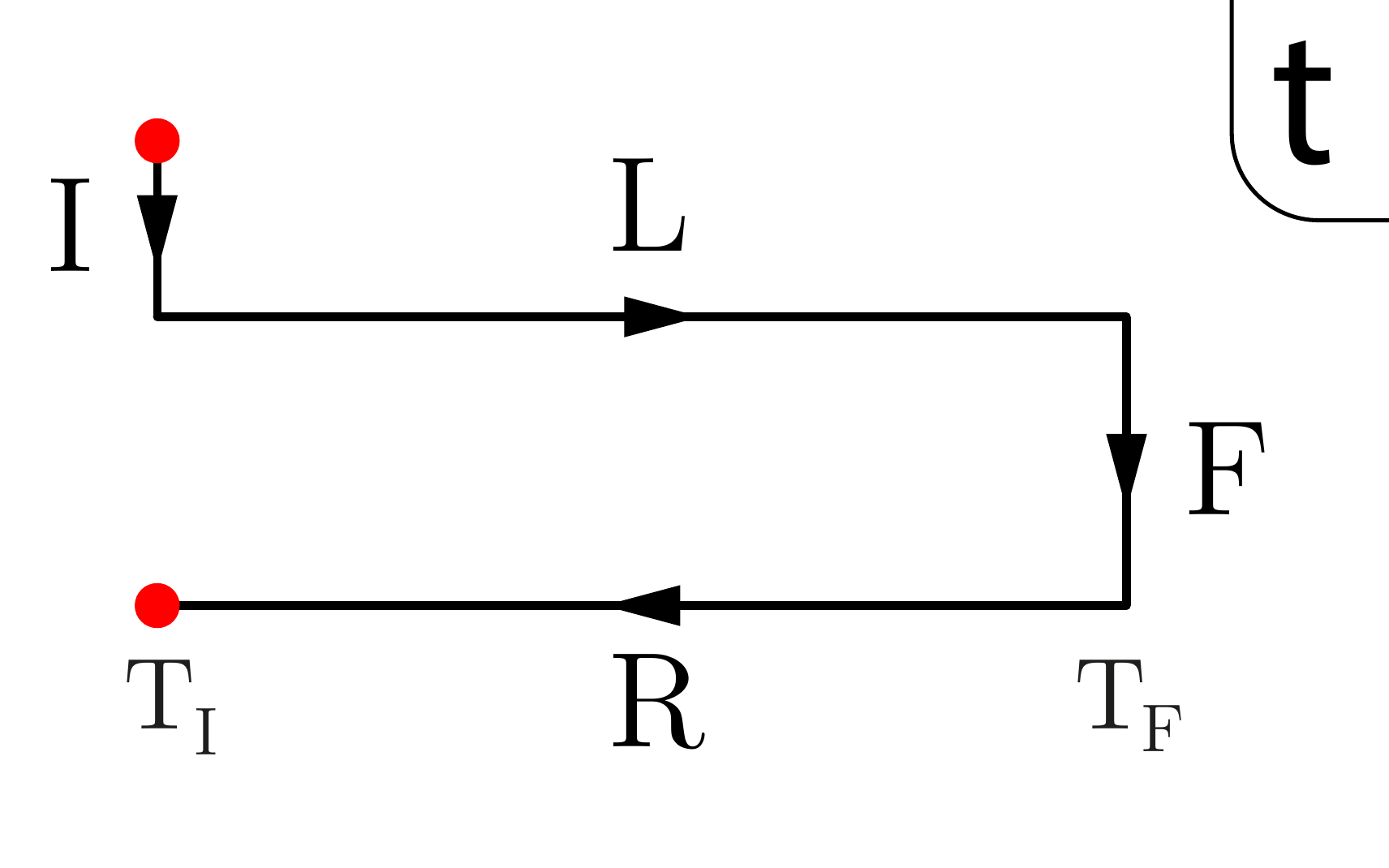}
\caption{}
\end{subfigure}
\begin{subfigure}{0.49\textwidth}\centering
\includegraphics[width=.9\linewidth] {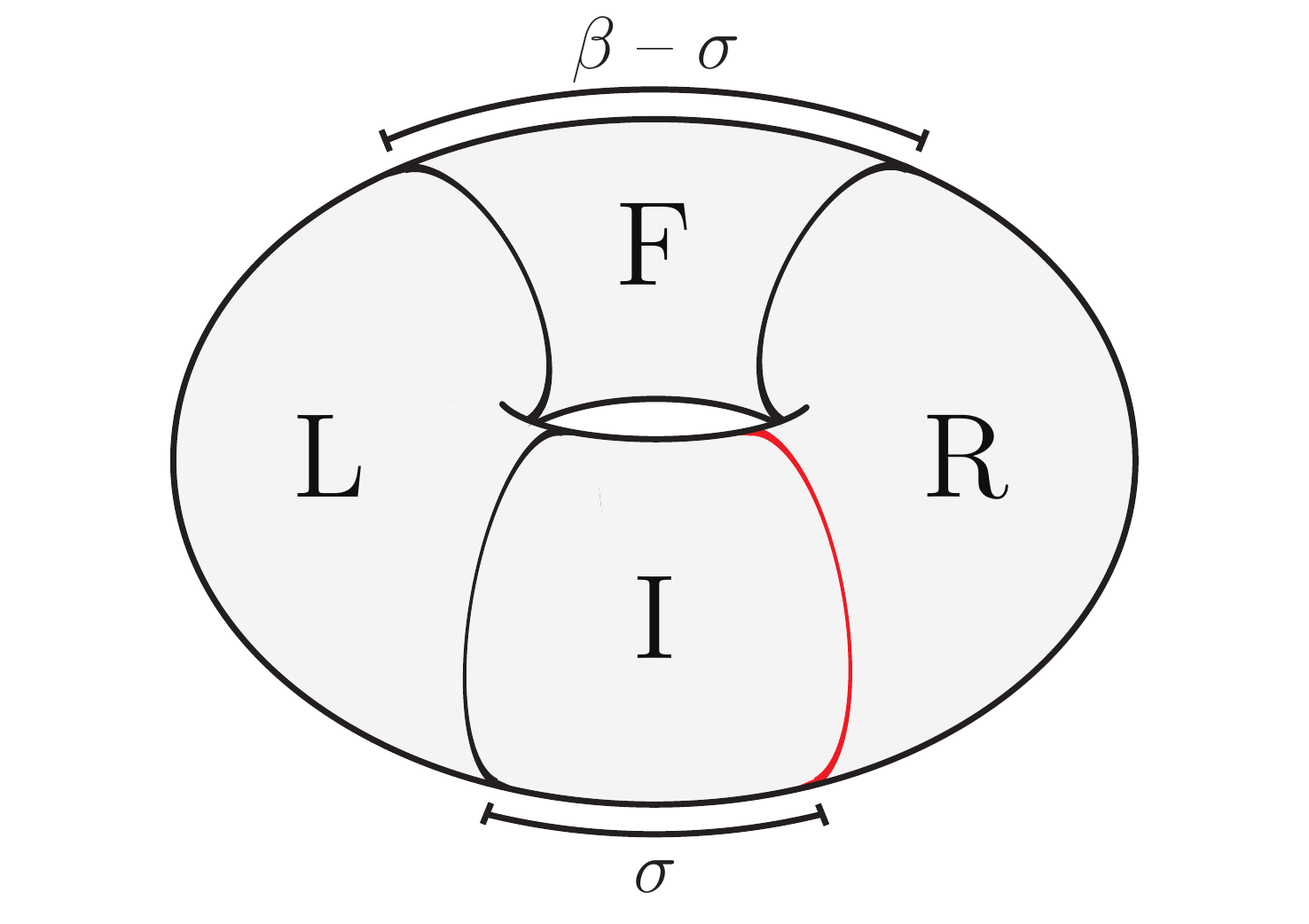}
\caption{}
\end{subfigure}
\caption{(a)  SK contour for Thermal AdS. The Euclidean pieces have different extensions: $\sigma$ for $I$ and $\beta-\sigma$ for $F$. (b) Dual real time Thermal AdS geometry. Notice that  the $L$ and $R$ regions are only connected through the Euclidean sections. The pieces are ordered as dictated by the contour order in fig. (a).}
\label{Fig:ThermalAdS}
\end{figure}

\subsection{Thermal AdS geometry}
\label{Thermal}

We summarize here the computations and results for the second bulk dual to the path in Fig. \ref{Fig:Camino}(a), i.e. Thermal AdS. 
The geometry is built by gluing together pure AdS cylinders resulting in a torus like solution as shown in Fig. \ref{Fig:ThermalAdS}(b). We carry over the notation above calling $L$ and $R$ the Lorentzian sections $t\in[-T/2,T/2]$. We will consider the Euclidean pieces to posses different temporal extensions: $\sigma$ for I and $\beta-\sigma$ for F as shown in Fig. \ref{Fig:ThermalAdS}(a). We will thus choose $\tau\in[\sigma-\beta,0]$ in F and $\tau\in[0,\sigma]$ in I, which for $\sigma=\beta/2$ reduces to the symmetric path of the previous section. It is important to stress that for the present geometry no physical connection appears between L and R regions.

The Lorentzian regions are defined for the AdS$_3$ metric
\begin{equation}\label{metric-AdS3}
ds^2=-(\tilde r^2+1)d \tilde t^2+\frac{d \tilde r^2}{\tilde r^2+1}+  \tilde r^2d\tilde \varphi^2\,,
\end{equation}
the ``tilde'' notation carried over from \eqref{metric-l1}, for $\varphi\sim\varphi+2\pi$ in both metrics. Inserting the ansatz $\Phi\propto e^{-i \omega \tilde t + i l \tilde \varphi} s(\omega,l,\tilde r)$ into the equation of motion \eqref{scalarEOM} gives
\begin{equation}\label{sNNAdS}
s(\omega,l,\tilde r) = \frac{\Gamma \left(\frac{1}{2} (|l|+\Delta -\omega )\right) \Gamma \left(\frac{1}{2} (|l|+\Delta +\omega )\right)}{\Gamma (\Delta -1) \Gamma (|l|+1)} \left(1+\tilde r^{2}\right)^{\omega/2}\,\, r^{|l|}\,\,
_2F_1\left( \frac{\omega+|l|+\Delta}{2}, \frac{\omega+|l|-\Delta+2}{2} ; 1+|l| ; -\tilde r^{2}\right)\,,
\end{equation}
as the only solution regular in the interior. The overall normalization is fixed, so that for generic $\{\omega,l\}$, $s(\omega,l,\tilde r) \approx   \tilde r^{\Delta-2}+\dots$ asymptotically. The  general solution on L is 
\begin{align}
\Phi_L(\tilde r,\tilde t,\tilde\varphi)= &\frac{1}{4\pi^2 } \sum_{l\in\mathbb{Z}} \int_{\mathcal{F}} d\omega e^{-i \omega \tilde t + i l \tilde\varphi}  \tilde \phi_L(\omega,l) s(\omega,l, \tilde r) 
+ \sum_{\substack{n\in\mathbb{N}\\ l\in\mathbb{Z} }}\Big(
 l_{nl}^{+}\, e^{- i \omega_{nl} \tilde t}+ l_{nl}^{-}\, e^{+ i \omega_{nl} \tilde t}\Big)e^{  i l \tilde \varphi}  s_{nl}(\tilde r)\,,
 \label{ThermalLSol}
\end{align}
with the $l_{nl}^{\pm}$ coefficients parametrizing the normalizable modes
\begin{equation}
\label{sNAdS}
s_{nl}(\tilde r)\equiv \oint_{\omega=-\omega_{nl}}d\omega \;s(\omega,l,\tilde r) \,,\qquad \omega_{nl}=2n+\Delta+|l|\;.
\end{equation}
These will become fixed once we glue the different pieces.  Equation \eqref{ThermalLSol} requires a choice of contour in the complex $\omega$-plane to avoid the singularities in the real axis depicted in Fig.\ref{Fig:Polos}(b). We refer the reader to \cite{SvRC} for details.  An analogous expression for the R region can be written. 

For the Euclidean regions normalizable modes suffice, then
\begin{align}
\Phi_I(\tilde r,\tilde \tau,\tilde \varphi)= 
\sum_{\substack{n\in\mathbb{N}\\ l\in\mathbb{Z} }}
 \left( i_{nl}^{+}\, e^{-  \omega_{nl} \tilde \tau }+i_{nl}^{-}\, e^{  \omega_{nl} \tilde \tau } \right) e^{ i l \tilde \varphi} s_{nl}(\tilde r)\,,
 \label{ThermalESol}
\end{align}
and similarly for region F. 
Following analogous steps as in \cite{SvRL,us}, the gluing conditions \eqref{bc} uniquely fix the coefficients $l^{\pm}_{nl}$, $i^{\pm}_{nl}$ as well as their R and F counterparts. 

The result of the gluing for a non-zero source on the L region, leads to
\be 
l^{\pm}_{nl}=\frac{1}{4\pi}\frac{\tilde\phi_L(\mp\omega_{nl},l)}{e^{\omega_{nl}\beta}-1}\;.
\label{coeL}
\ee
By inserting \eqref{coeL} back into \eqref{ThermalLSol}, one can rewrite the bulk field as
\begin{align}
\Phi_L(\tilde r,\tilde t,\tilde \varphi)= &\frac{1}{4\pi^2 } \sum_{l\in\mathbb{Z}} \int_{\mathcal{R}} d\omega d\tilde t' d\tilde \varphi' e^{-i \omega (\tilde t-\tilde t') + i l (\tilde \varphi-\tilde \varphi')} \frac{e^{\omega\beta}}{e^{\omega\beta}-1} \phi_L(\tilde t',\tilde \varphi') s(\omega,l,\tilde r)  \nn\\
&-
\frac{1}{4\pi^2 } \sum_{l\in\mathbb{Z}} \int_{\mathcal{A}} d\omega d\tilde t' d\tilde \varphi' e^{-i \omega (\tilde t-\tilde t') + i l (\tilde \varphi-\tilde \varphi')} \frac{1}{e^{\omega\beta}-1} \phi_L(\tilde t',\tilde \varphi') s(\omega,l,\tilde r) 
\label{ThermalLSol2}
\end{align}
where $\mathcal{R}$ and $\mathcal{A}$ stand for the retarded and advanced integrations paths respectively. Notice that \eqref{ThermalLSol2} mimics the structure of \eqref{Lsol} when using \eqref{AandB}.

\section{Bulk Correlators}\label{Correlators}

We will now obtain the 2-pt large $N$  CFT correlators from the field solutions obtained in the previous section following prescription \eqref{SvRPath} and compare our results with CFT predictions. We build the complete bulk action as a sum over four pieces 
$$i S_{\cal C}=-S_{I} + iS_{L} - S_{F} + iS_{R}\;.$$
The saddle point approximation evaluates each action on-shell, leaving only boundary terms for each piece. 
Since Euclidean sources have been turned off contributions from I and II vanish, then
\begin{align}\label{Onchell}
iS_{\cal C}^0\left[\phi_L,\phi_R\right]&=-\,\frac i2 \int_{\cal C}  \sqrt{\gamma} \;\Phi\; n^\mu \partial_\mu \Phi 
=-\,\frac{i}{2}  r^{\Delta} \left[\int_0^T dt d\varphi\;\phi_L \left(r\partial_r\Phi_L\right)
-\int_0^T dt d\varphi\;\phi_R \left(r\partial_r\Phi_R\right)
\right]_{r\to\infty}\,.
\end{align}
Evaluating \eqref{Onchell} on metrics \eqref{metric-l2} and \eqref{metric-AdS3} provide the real time propagator matrices in the high and low temperatures regimes of the dual CFT. The sign difference between the terms arises from the choice of time parametrization, see Fig. \ref{Fig:Boots}(b).

We will first concentrate on the diagonal elements of the matrix \eqref{PropMatrix-CFT}. For the BH geometry depicted in Fig. \ref{Fig:Camino} (b) one obtains
\begin{align}\label{G11BH}
\langle\!\langle\Psi_0|{\cal O}_L (t,\varphi) {\cal O}_L ( t', \varphi')|\Psi_0\rangle\!\rangle \Big|_{BH}
&\equiv -i \frac{\delta^2 S_{\cal C}^0}{\delta \phi^L\delta \phi^L} \nn\\
&= \frac{(\Delta-1)}{2\pi^2 i r_S} \sum_{l} \int d\omega e^{-i\omega \Delta t+i l \Delta \varphi} \left( \frac{-1}{e^{2\pi\omega}-1} \alpha_{\omega l\Delta}\beta_{\omega l\Delta} + \frac{e^{2\pi\omega} }{e^{2\pi\omega}-1} \alpha_{(-\omega) l\Delta}\beta_{(-\omega) l\Delta}\right)\nn\\ 
&=\sum_{j\in\mathbb{Z}}\frac{(\Delta-1)^2}{2^{\Delta-1}\pi}\left[\cosh(\Delta \varphi+2\pi r_S j)-\cosh(\Delta t(1-i\epsilon))  \right]^{-\Delta} 
\end{align}
We have relegated the momentum integration to App. \ref{App:Prop}. Some comments on \eqref{G11BH} regarding causality are in order:  the pole structure of $\beta_{\pm\omega l \Delta}$ guarantees that the second line in \eqref{G11BH} appropriately reproduces the expected advanced/retarded propagators shown in \eqref{SKprop}. The result is in agreement with \cite{herzog,SvRL} and, in line with the holographic real time viewpoint, the sign of the $i\epsilon$ regulator is fixed by convergence. The propagator \eqref{G11BH} can be seen to meet the KMS condition, i.e. it is invariant under $\Delta t\to \Delta t+i\beta$. As stressed in \cite{eternal}, it vanishes exponentially as $\Delta t\to \infty$, manifesting that, at high temperature, correlations are quickly lost.

The result for Thermal AdS follows analogously from \eqref{ThermalLSol} and \eqref{Onchell}, the result is
\begin{align}\label{G11T}
\langle\!\langle\Psi_0|{\cal O}_L (  t,  \varphi) {\cal O}_L (  t',  \varphi')|\Psi_0\rangle\!\rangle\Big|_{Th} 
&=\sum_{j\in\mathbb{Z}}\frac{(\Delta-1)^2}{2^{\Delta-1}\pi}\left[\cos(\Delta  t(1-i\epsilon)+i \beta j) -\cos(\Delta  \varphi) \right]^{-\Delta}\;.
\end{align}
Again, from its momentum expansion, which can be read off from \eqref{ThermalLSol}, one finds the expected structure shown in \eqref{SKprop}. The result can be understood as a sum over $t$-images, with period $\beta$, of the zero temperature 2-pt function, thus ensuring the KMS condition. Propagation of information in this solution, in contrast with \eqref{G11BH}, is  oscillatory as expected at low temperature.
Finally, expressions \eqref{G11BH} and \eqref{G11T} which respectively dominate in the $\beta\lessgtr\beta_c$\footnote{ Recall that $\beta_c=2\pi R_{AdS}=2\pi$ in our units.}  describe the CFT dynamics for any temperature. 

Computations for correlators in the R region easily follow using the solutions analogous to the ones found in the previous section. The result is, 
\begin{align}\label{G22T}
\langle\!\langle\Psi_0|{\cal O}_R (t,\varphi) {\cal O}_R ( t', \varphi')|\Psi_0\rangle\!\rangle\Big|_{BH} &\equiv -i \frac{\delta^2 S_{\cal C}^0}{\delta \phi^R\delta \phi^R} \nn\\
&=\sum_{j\in\mathbb{Z}}\frac{(\Delta-1)^2}{2^{\Delta-1}\pi}\left[\cosh(\Delta \varphi+2\pi r_S j)-\cosh(\Delta t(1+i\epsilon))  \right]^{-\Delta}\; . 
\end{align}
It is reassuring to see that the answer is consistent with the following TFD property: the second diagonal element in \eqref{PropMatrix-CFT}  is the reverse time-order propagator, which can be obtained as the complex conjugate of the first diagonal element. 
Similar arguments hold for the low temperature propagator, which can be obtained as the complex conjugate of \eqref{G11T}. Notice that the diagonal elements of the matrix in the high and low temperature regime, e.g.  \eqref{G11BH} and \eqref{G11T},   are related via the standard double Wick rotation $\phi\to it$, $t\to i \phi$. 

An interesting analysis comes from the off-diagonal elements of the propagator matrix. These element require field solution with sources on L and R turned on. As observed in \cite{herzog}, large $N$ holographic computations using \eqref{SvRPath} give symmetric propagator matrices,
\begin{align}\label{G12}
\langle\!\langle\Psi_0|{\cal O}_L (t,\varphi) {\cal O}_R ( t', \varphi')|\Psi_0\rangle\!\rangle \equiv -i \frac{\delta^2 S_{\cal C}^0}{\delta \phi^L \delta \phi^R}=-i \frac{\delta^2 S_{\cal C}^0}{\delta \phi^R \delta \phi^L} 
 \equiv \langle\!\langle\Psi_0|{\cal O}_R (t',\varphi') {\cal O}_L ( t, \varphi)|\Psi_0\rangle\!\rangle\;,
\end{align}
whilst CFT computations distinguishes between different $\sigma\in[0,\beta)$ SK paths shown in Fig. \ref{Fig:ThermalAdS}(a). 
We can holographically compute the general $\sigma$ off-diagonal elements of the propagator matrix in the Thermal scenario, yielding
\begin{align}\label{ThermalG12}
\langle\!\langle\Psi_0|{\cal O}_R ( t', \varphi') {\cal O}_L (t, \varphi ) |\Psi_0\rangle\!\rangle\Big|_{Th} \equiv& -i \frac{\delta^2 S_{\cal C}^0}{\delta \phi^L\delta\phi^R}\nn\\
=&\frac 12\sum_{j\in\mathbb{Z}}\frac{(\Delta-1)^2}{2^{\Delta-1}\pi}\left[\cos(\Delta  t+i\beta j-i\sigma) -\cos(\Delta  \varphi) \right]^{-\Delta}\nn\\
&+\frac 12\sum_{j\in\mathbb{Z}}\frac{(\Delta-1)^2}{2^{\Delta-1}\pi}\left[\cos(\Delta  t+i\beta j-i(\beta-\sigma)) -\cos(\Delta \varphi) \right]^{-\Delta}\;.
\end{align}
This expression arises from two contributions, which can be  understood as the two possible paths connecting the points in  L and R segments through the $\sigma$ and $\beta-\sigma$ Euclidean sections. 
The images sum accounts for the windings of these paths. Each of the inequivalent contributions are in fact what is expected for the off-diagonal elements of the $\sigma\neq\beta/2$ propagator matrix. One can see that the semi-classical limit is taking the mean of the two quantum behaviors, although the shorter path dominates exponentially over the longest one. The contributions add up in the $\sigma=\beta/2$ case, recovering the symmetric matrix \eqref{PropMatrix-CFT}. One should note that no $i\epsilon$ regulator is needed for convergence in this case\footnote{Except for the  $\sigma=0$ or $\sigma=\beta$ cases  which were considered in \cite{SvRL}}.

The black hole geometry, dual to the $\sigma=\beta/2$ path gives
\begin{align}
\langle\!\langle\Psi_0|{\cal O}_R (t,\varphi) {\cal O}_L ( t', \varphi')|\Psi_0\rangle\!\rangle\Big|_{BH} &\equiv -i \frac{\delta^2 S_C^0}{\delta \phi^L\delta\phi^R}\nn\\
&= \frac{2(\Delta-1)}{4\pi^2 i r_S} \sum_{l} \int d\omega \; e^{-i\omega \Delta t+i l \Delta \varphi} \frac{e^{\pi\omega} }{e^{2\pi\omega}-1}\left( -  \alpha_{\omega l\Delta}\beta_{\omega l\Delta} +  \alpha_{(-\omega) l\Delta}\beta_{(-\omega) l\Delta}\right)\nn\\
 &=\sum_{j\in\mathbb{Z}}\frac{(\Delta-1)^2}{2^{\Delta-1}\pi}\left[\cosh(\Delta \varphi+2\pi r_S j)+\cosh(\Delta t)  \right]^{-\Delta}\;,\label{BHG12}
\end{align}
where the sign difference in the second line agrees with \eqref{offdiag} and the third line matches \cite{eternal}.

We now make some comments regarding \eqref{ThermalG12} and \eqref{BHG12}. First, notice that in contrast to the case of diagonal matrix elements, one cannot relate the results via the double Wick rotation $\phi\to it$, $t\to i \phi$. This stems from the $\sigma$ terms in \eqref{ThermalG12} which do not rotate. The result is related to the topology of the solutions: unlike the Thermal solution, the black hole admits paths connecting L and R through the wormhole, cf. Figs. \ref{Fig:Camino}(b) and \ref{Fig:ThermalAdS}(b).
The fact of  \eqref{ThermalG12} and \eqref{BHG12} not being connected by a double Wick rotation manifest the importance of understanding  the correspondence in the a real-time setup. 

One can also interpret these correlators as providing information on the entanglement between the L and R dof's \cite{vanram}. Results \eqref{ThermalG12} and \eqref{BHG12},  understood as lower bounds on mutual information, behave as expected in disentangling experiments.

\section{Conclusions}
\label{Conclusions}

In this work we constructed a novel geometry dual to the symmetric SK path presented in Fig. \ref{Fig:Camino}.
The solution was obtained by gluing Euclidean and Lorentzian sections of the AdS black hole and represents the gravity dual to the real-time extension of a finite temperature CFT in the high temperature limit.
The particular SK path chosen shows several appealing properties that we explored.

\vspace{2mm}

We would like to mention the following remarkable properties of the  solution:

1. It removes the black hole interiors and therefore the singularities, as it happens for the Euclidean AdS BH, but its Lorentzian region remains two-sided, i.e with R- and L-wedges. In this sense, it is a natural real time evolution of the TFD state built from the Euclidean pieces.

2. The SK path with two equal $\beta/2$ lengths allows to formulate the problem in accordance with the TFD approach \cite{TFDUme}. Our solution 
gives the dual description of the TFD evolution. 
In other words, the gravitational dual of the operator $\mathbb{U}$, defined in \eqref{UTFD}, is represented by the two sided BH under the boost time evolution as shown in Fig. \ref{Fig:Boots}. Indeed, the Hartle-Hawking-Maldacena wave function given by the Euclidean semi-disk geometry, remains invariant throughout its Lorentzian evolution.

3. The Euclidean regions on their own have also a physical interpretation, providing initial and final states in the TFD viewpoint. This matches the observation that the HHM-state is the TFD thermal vacuum \cite{eternal}, whose gravity dual is precisely identified with the regions I/F of the actual geometry.

4. The complete manifold is made from three spacetimes and covers the four pieces of the path in Fig. \ref{Fig:Camino}(a): $L\cup R$ belongs to a single black hole. This is in line with the intuition that high temperature regime increasingly entangling the system, made holographic by connecting the regions through the wormhole \cite{vanram,MarceEssay}.

5. For the chosen SK path, the analyticity of the fields in the bulk interior is forced by the gluing conditions \eqref{bc}. This is relevant for the Lorentzian regions since the field configurations end up being analytic through the wormhole. In this sense, our solution geometrically captures the Unruh trick. The traditional analytic extension required to obtain global positive energy modes on a $L\cup R$ spacelike surface, is automatically incorporated by the Euclidean regions $I/F$. 

6. The solution, together with the real time Thermal AdS (depicted in Fig.\ref{Fig:ThermalAdS}) extend the two saddle points Hawking-Page scenario to real-time. We have explored the transition studying 2-pt functions in the two different regimes. We have recovered the propagator matrix and checked the path ordering in the CFT.

7. Based on analyticity arguments, it does not seem to exist an analogous BH solution for SK paths other than the symmetric one. While the Thermal scenario admits arbitrary Euclidean path lengths, 
the $\sigma=\beta/2$ path is privileged in the sense that the CFT matrix propagator becomes symmetric
which is always the case from the bulk viewpoint, at least in the semi-classical limit.

8. We found that the off-diagonal elements of the propagator matrix in the high and low temperature regimes are not connected through the double Wick rotation which connects the metrics \eqref{metric-l2} and \eqref{metric-AdS3}. The result is related to the topology of the solutions, in the BH the Lorentzian regions are connected through the wormhole. This represents a non-trivial example stressing the importance of a better real-time understanding of the correspondence.

9. The off-diagonal elements in the matrix propagator provide information on the entanglement of the dof's and the classical connectivity between the two sides of the spacetime. 

As mentioned in the main text, the immediate outlook from this work is to study holographic excited states in our solution by turning on Euclidean sources \cite{us,us2}. Since we have been able to reinterpret the finite temperature problem as a scattering process in a thermal bath,  we expect the excitations to be associated to (thermal) coherent states \cite{EssayBoots}. It will be interesting however, to study how this prescription defines the nature of the new coherent states in terms of the normal modes in the duplicated TFD Hilbert space. We would also like to study whether this excited states scenario favors, in a similar fashion as the comments in 7., the symmetric $\sigma=\beta/2$ SK path. We will pursue this objectives in \cite{tocome}. 

It would be also interesting to study backreaction in some perturbative set up, upon imposing boundary sources that excite the TFD ground state, and study how the present geometry would eventually be deformed \cite{Marolf2}. 
One might also consider gluing multiple copies of the Lorentzian wedges for OTOC computations \cite{OTOC}.
Generalizations of our geometry that include charge and angular momentum would also be of interest.
Another line of future research would be deforming the TFD (decoupled) field theory Hamiltonian $H_R - H_L$, with a local coupling term $\sim g \;{\cal O}_R\, {\cal O}_L$ leading to a traversable wormhole 
\cite{Raulo,Pingao,Malda1+1,BootsBrazil,MarceEssay}. 

~

\nin {\bf Acknowledgements}

\nin  Work supported by  UNLP and CONICET grants X791, PIP 2017-1109 and PUE B\'usqueda de nueva F\'\i sica.

\appendix

\section{Appendix: Propagator in configuration space} \label{App:Prop}
In this appendix we show that 
\begin{align}
\langle{\cal O}_L (t,\varphi) {\cal O}_L ( t',  \varphi')\rangle &\equiv -i \frac{\delta^2 S_{\cal C}^0}{\delta \phi^L\delta \phi^L}\nn\\
&= \frac{2(\Delta-1)}{4\pi^2 i r_S} \sum_{l} \int d\omega e^{-i\omega (t-  t')+i l (\varphi- \varphi')}\left( \frac{-1}{e^{2\pi\omega}-1} \alpha_{\omega l \Delta}\beta_{\omega l \Delta} + \frac{e^{2\pi\omega} }{e^{2\pi\omega}-1} \alpha_{(-\omega) l \Delta}\beta_{(-\omega) l \Delta} \right) \nn\\
&= \frac{(\Delta-1)}{2\pi^2 i r_S} \sum_{l} \int d\omega e^{-i\omega \Delta t+i l \Delta\varphi}\left( - n_\omega \alpha_{\omega l \Delta}\beta_{\omega l \Delta} + (1+n_\omega) \alpha_{(-\omega) l \Delta}\beta_{(-\omega) l \Delta} \right) \label{A1}\\
&=\sum_{j\in\mathbb{Z}}\frac{(\Delta-1)^2}{2^{\Delta-1}\pi}\left[\cosh(\Delta \varphi+2\pi r_S j)-\cosh(\Delta t(1-i\epsilon))  \right]^{-\Delta}\label{A2}
\end{align}
where $n_\omega\equiv(e^{2\pi\omega}-1)^{-1}$, $\alpha_{\omega l \Delta}$ and $\beta_{\omega l \Delta}$ coefficients where defined in \eqref{alpha} and \eqref{beta} and the Feynman regulator $\epsilon>0$ is mandatory for convergence. The black hole mass has been transferred to the periodicity of the polar angle $\varphi\sim\varphi+2\pi r_S$, therefore, the $l$-summation is over $l=k/r_S$ with $k\in\mathbb Z$, see \eqref{rescaling}.

One comment regarding the normalization of 2-pt functions is at hand. The factor $2(\Delta-d)=2(\Delta-1)$ in the numerator of the second line onwards does not follow directly from the field solutions found in \eqref{Lsol}, which yields $2\Delta$. The correction arises from the different ways of regularizing the divergences arising in the asymptotic boundary. This is however not related to our concrete problem and has already been extensively covered in the literature, see e.g. \cite{us2,Rastelli,HolRenorm}. We emphasize that the correct coefficient from an AdS/CFT perspective is the one kept in \eqref{A2}. 

In order to do this, we will pick a space-like distance, find the correct $\epsilon>0$ needed for convergence near the contact points. Take $\Delta \phi>\Delta t>0$ such that $\Delta \phi+\Delta t>0$ and $\Delta \phi-\Delta t>0$. Analyticity forces the result everywhere else in the $\{t,\phi\}$ plane away from the light-cones. We are left with the task to infer the $i\epsilon$ prescription implicit in \eqref{A1}. This will be done taking $\Delta \phi=0$ and $0<\Delta t\ll 1$ and $-1\gg\Delta t<0$ in turn to see that the Feynman prescription, i.e. \eqref{A2} is correct. 

The $\omega$-integral in \eqref{A1} will be performed by making use of the residue theorem, so we need to know the pole structure of the integrand:

$\bullet$ $\alpha_{\omega l \Delta}$ is a polynomial in $\omega$ and therefore analytic in the whole complex-plane. 

$\bullet$ $n_\omega$ has poles at $\omega=i s$, $s\in\mathbb{Z}$. However, the product $(n_\omega \alpha_{\omega l \Delta}\beta_{\omega l \Delta})$ is regular at $\omega=0$.

$\bullet$ $\beta_{\omega l \Delta}$ has poles at $\omega=\pm l+i(2n+\Delta),~ n\in\mathbb N$, i.e. the upper half plane. 

The proof of \eqref{A2} will be split in 3 parts: first, show that the only poles that contribute to the 2-pt function are the ones arising from $\beta_{\omega l \Delta}$ and $\beta_{(-\omega) l \Delta}$.
Second, we get from \eqref{A1} to the expression \eqref{A2} in a space-like frame. Thirdly we show that $\Delta t(1-i\epsilon)$ is the mandatory regulator starting from \eqref{A1}.
Every other momentum integral in our BH solution follows, albeit details, from this one.

\vspace{.2cm}

\nin {\sf 1.  Residues from $n_\omega$ do not contribute}.

\vspace{.2cm}

Consider
\begin{align}
{\cal I}(\Delta t,\Delta\varphi) &\equiv \frac{ (\Delta -1) }{2 \pi ^2 i r_S} \sum_{l} \int d\omega e^{-i\omega \Delta t+i l \Delta \varphi} \left( - n_\omega \alpha_{\omega l \Delta}\beta_{\omega l \Delta} + \left(1+n_\omega \right) \alpha_{(-\omega) l \Delta}\beta_{(-\omega) l \Delta}\right) \label{A3}\\
&=\frac{ (\Delta -1) }{2 \pi ^2 i r_S} \sum_{l\in\mathbb Z} \int d\omega e^{-i\omega \Delta t+i \frac {l}{r_S} \Delta \varphi} \left( - n_\omega \alpha_{\omega \frac {l}{r_S} \Delta}\beta_{\omega \frac {l}{r_S} \Delta} + \left(1+n_\omega \right) \alpha_{(-\omega) \frac {l}{r_S} \Delta}\beta_{(-\omega) \frac {l}{r_S} \Delta}\right)\;,\nn
\end{align}
where we made explicit the $r_S$ factor in the angular momentum sum, see \eqref{rescaling}. To explicitly compute the integral we choose $\Delta t>0$. This demands to close the $\omega$-integral in the lower half complex $\omega$-plane. In the following we will only consider the contributions arising from the residues of $n_\omega$. Closing the contour through the lower half-plane one find
\begin{align}
{\cal I}(\Delta t,\Delta\varphi)  
   &=\frac{(\Delta -1) }{2 \pi ^2 r_S}\sum _{l\in\mathbb Z} \sum _{n=1} e^{-n\Delta t+i \frac l{r_S} \Delta \phi } \left(\alpha_{(-in) \frac {l}{r_S} \Delta} \beta_{(-in) \frac {l}{r_S} \Delta} - \alpha_{(in) \frac {l}{r_S} \Delta} \beta_{(in) \frac {l}{r_S} \Delta}\right)+\left(\beta\text{ residues}\right)\nn\;.
\end{align}
Making use of the Poisson re-summation trick we can translate the sum over $l$ into an integral and a sum over images\footnote{The convention for Fourier transformation are: $\tilde f(k)=\int dx\,f(x)e^{-i2\pi kx}$ and $f(x)=\int dk\,\tilde f(x)e^{i2\pi kx}$ }  
\begin{equation}
\label{PoissonRessumation}
\sum _{l\in Z} f\left(\frac l a\right) = a \sum _{m\in \mathbb Z} \tilde f(m a)\qquad\Longrightarrow \qquad \sum _{l\in\mathbb Z} f\left(\frac l {r_S}\right)= r_S\sum _{m\in\mathbb Z} \int \;dl\; f(l) \;e^{i (2 \pi r_S m) l} 
\end{equation}
Inserting it into the above expression yields 
\begin{align}
{\cal I}(\Delta t,\Delta\varphi)  
   &=\frac{(\Delta -1) }{2 \pi ^2 }\sum _{n=1} \sum _{m\in Z} \int \;dl \; e^{-n \Delta t +i l ( \Delta \phi + 2 \pi r_S m )} \left(\alpha_{(-in) l \Delta} \beta_{(-in) l \Delta} - \alpha_{(in) l \Delta} \beta_{(in) l \Delta}\right)+\left(\beta\text{ residues}\right)\nn,
\end{align}

now, one can see that the integrand between parenthesis, for integer $\Delta\geq2$ and $n\geq1$, becomes $l$ polynomials which have no poles: each terms has poles that cancel each other among themselves. The $l$ integral now can be closed either in the upper or lower half plane due to the exponential, so the lack of poles means that these terms do not contribute. The $\Delta t<0$ case follows the same way.

\vspace{.2cm}

\nin {\sf 2. Residues from $\beta_{\omega l \Delta}$}.

\vspace{.2cm}

We now consider the contributions from the poles in $\beta_{\omega l \Delta}$. From \eqref{A3},
\begin{equation}\nn
{\cal I}(\Delta t,\Delta\varphi) = \frac{ (\Delta -1) }{2 \pi ^2 i r_S} \sum_{l} \int d\omega e^{-i\omega \Delta t+i l \Delta \varphi} \left( - n_\omega \alpha_{\omega l \Delta}\beta_{\omega l \Delta} + \left(1+n_\omega \right) \alpha_{(-\omega) l \Delta}\beta_{(-\omega) l \Delta}\right)\;
\end{equation}
we consider $\Delta t>0$, close downwards and pay attention to the poles $\omega=\pm l -i(2s+\Delta)$, $s\geq 0$ coming from $\beta_{(-\omega) l \Delta}$ leading to
\begin{align}\nn
{\cal I}(\Delta t,\Delta\varphi) =\frac{2 (-1)^{\Delta -1}}{\pi i r_S  \Gamma (\Delta -1)^2}& \sum_{s\geq0} (-s-\Delta
   +1)_{\Delta -1} e^{-(2s+\Delta) \Delta t}  \sum_{l}  \sum_{\pm} \; e^{i l (\Delta \varphi \mp \Delta t)}\left(1+n_{\pm l}\right) (\mp i l-s-\Delta
   +1)_{\Delta -1}\;,
\end{align}
where we have used that $n_{\pm l-i(2s+\Delta)}=n_{\pm l}$ for $(2s+\Delta)\in\mathbb{Z}$ and the explicit definition of $\alpha_{(-\omega) l \Delta}$ \eqref{alpha}. We remark that the $l=0$ term is regular for the $n_{0}$ singularity cancels. We focus now on the sum on $l$, again by Poisson re-summation \eqref{PoissonRessumation},
\begin{align}
{\cal M} &\equiv \sum_{l} \sum_{\pm} \; e^{i l (\Delta \varphi \mp \Delta t)}\left(1+n_{\pm l}\right) (\mp i l-s-\Delta +1)_{\Delta -1}\label{A5}\\
&=\sum_{l}\; e^{i l (\Delta \varphi - \Delta t)}\left(1+n_l\right) (- i l-s-\Delta +1)_{\Delta -1}+e^{i l (\Delta \varphi + \Delta t)}\left(1+n_{- l}\right) (+ i l - s - \Delta +1)_{\Delta -1}\; \nn\\
&=r_S\sum_{m\in\mathbb{Z}} \int dl \;e^{i(2\pi r_S m) l}\;\left(e^{i l (\Delta \varphi - \Delta t)}\left(1+n_l\right) (- i l-s-\Delta +1)_{\Delta -1}+e^{i l (\Delta \varphi + \Delta t)}\left(1+n_{-l}\right) (+ i l-s-\Delta +1)_{\Delta -1}\right)\; \nn.
\end{align}
We now pick the left spacelike (lightcone) quadrant where $\Delta \phi\pm\Delta t>0$, such that both $l$ integrals close downwards, giving
\begin{align}
{\cal M} &\equiv i r_S\sum_{m\in\mathbb{Z}}  \sum_{j=1}^\infty \;e^{i(2\pi r_S m) l}\;\left(e^{-j (\Delta \varphi - \Delta t)} ( + j -s-\Delta +1)_{\Delta -1}-e^{-j (\Delta \varphi + \Delta t)} (- j -s-\Delta +1)_{\Delta -1}\; \right)\nn\\
&= i r_S\sum_{m\in\mathbb{Z}} \sum_{j=1}^\infty \sum_{\pm} \;(\pm1) e^{i(2\pi r_S m) l} e^{-j (\Delta \varphi \mp \Delta t)} ( \pm j -s-\Delta +1)_{\Delta -1} \nn
\end{align}
so that \eqref{A1} becomes
\begin{align}\nn
\frac{2 (-1)^{\Delta -1}}{\pi \Gamma (\Delta -1)^2}& \sum_{s\geq0} (-s-\Delta
   +1)_{\Delta -1} e^{-(2s+\Delta) \Delta t}  \sum_{m\in\mathbb{Z}}  \sum_{j=1}^\infty \sum_{\pm} \;(\pm1) e^{i(2\pi r_S m) l} e^{-j (\Delta \varphi \mp \Delta t)} ( \pm j -s-\Delta +1)_{\Delta -1}
\end{align}
which can be summed for $2<\Delta\in\mathbb{Z}$ and extended for general values giving
\begin{align}\nn
\sum_{m\in\mathbb{Z}}\frac{(\Delta-1)^2}{2^{\Delta-1}\pi}[-\cosh(\Delta t)+\cosh(\Delta \phi+2\pi r_S m)]^{-\Delta}\,.
\end{align}
This result extends by analytic extension to other points outside of the lightcone. We now show that \eqref{A1} is so that it forces the Feynman regulator.

\vspace{.2cm}

\nin {\sf 3. Feynman Regulator}.

\vspace{.2cm}

To uncover the regulator we go back to \eqref{A5}
\begin{align}
{\cal M}= \sum_{l} \sum_{\pm} \; e^{i l (\Delta \varphi \mp \Delta t)}\left(1+n_{\pm l}\right) (\mp i l-s-\Delta +1)_{\Delta -1}\nn
\end{align}
and take the limiting case $\Delta \varphi=0$ and $\Delta t(1-i\epsilon)\sim -i\epsilon$ recalling that we needed $\Delta t>0$ to get there. We now have 
\begin{align}
{\cal M}=\sum_{l} & \; e^{-l\epsilon}\left(1+n_l\right) (- i l-s-\Delta +1)_{\Delta -1}+e^{+l\epsilon}\left(1+n_{-l}\right) (+ i l-s-\Delta +1)_{\Delta -1}\;\nn,
\end{align}
where each term on its own is well behaved in the $l\to\pm\infty$ limits. The Pochhammer symbols are polynomials while $e^{\mp l\epsilon}(1+n_{\pm l})$ are exponentially convergent. The analogous result for $\Delta t<0$, yields
\begin{align}
{\cal M}' =\sum_{l} \sum_{\pm}\; e^{i l (\Delta \varphi \mp \Delta t)} n_{\pm l} (\pm i l-s-\Delta +1)_{\Delta -1}\;\nn
\end{align}
but now $\Delta t(1-i\epsilon)\sim +i\epsilon$ leading to 
\begin{align}
{\cal M}' =\sum_{l} \; e^{ l\epsilon} n_l (+ i l-s-\Delta +1)_{\Delta -1}+e^{ -l\epsilon} n_{- l} (- i l-s-\Delta +1)_{\Delta -1}\;\nn 
\end{align}
which again contains the correct regulator for each separate term. This completes the demonstration of \eqref{A2}. Our result agrees with \cite{SvRL}.


\end{document}